\documentclass[12pt]{article}
\usepackage{epsfig}
\newcommand{\nn}{\nonumber}
\newcommand{\raw}{\rightarrow}

\newcommand{\be}{\begin{equation}}
\newcommand{\ee}{\end{equation}}
\newcommand{\bea}{\begin{eqnarray}}
\newcommand{\eea}{\end{eqnarray}}

\newcommand{\dd}{\displaystyle}

\newcommand{\PCPV}{
\begin{picture}(22,10)
\put(8,-2){\line(2,1){12}}
\put(0,0){$P_{CP}$}
\end{picture}}
\newcommand{\PCPC}{
\begin{picture}(22,10)
\put(0,0){$P_{CP}$}
\end{picture}}
\begin{document}
%
%
\thispagestyle{empty}
\begin{flushright}
{\tt hep-ph/9909254}\\
{CERN-TH/99-252}\\
{FTUAM-99-25}\\
{FTUV/99-59}\\
{IFIC/99-62}
\end{flushright}
\vspace*{1cm}
\begin{center}
{\Large{\bf Neutrino mixing and CP-violation} }\\
\vspace{.5cm}
A. Donini$^{\rm a,}$\footnote{donini@daniel.ft.uam.es},
M.B. Gavela$^{\rm a,}$\footnote{gavela@garuda.ft.uam.es},
P. Hern\'andez$^{\rm b,}$\footnote{pilar.hernandez@cern.ch. On leave 
from Dept. de F\'{\i}sica Te\'orica, Universidad de Valencia.}
and S. Rigolin$^{\rm a,}$\footnote{rigolin@daniel.ft.uam.es}
 
\vspace*{1cm}
$^{\rm a}$ Dept. de F\'{\i}sica Te\'orica, Univ. Aut\'onoma de
Madrid, 28049 Spain \\
$^{\rm b}$ Theory Division, CERN, 1211 Geneva 23, Switzerland

\end{center}
\vspace{.3cm}
\begin{abstract}
\noindent
The prospects of measuring the leptonic angles and CP-odd phases at a 
neutrino factory are discussed in two scenarios: 
1) three active neutrinos as indicated by the present ensemble of 
atmospheric plus solar data; 
2) three active plus one sterile neutrino when the LSND signal is also 
taken into account. For the latter we develop one and two mass dominance 
approximations. The appearance of wrong sign muons in long baseline 
experiments and tau leptons  in short baseline ones provides the best tests 
of CP-violation in scenarios 1) and 2), respectively.
\end{abstract}

\newpage

%
%
\section{Introduction}
%
%
The putative value of the masses and mixing angles
of the leptonic sector are part of the fundamental problem usually dubbed 
``flavour puzzle''. The comprehension of the origin of these 
parameters and the analogous ones in the quark sector, together 
with the overall understanding of the origin of masses 
attempted through Higgs searches, constitutes one of the most
passionating research subjects today in particle physics. 
The precise determination of those parameters is a mandatory 
first step.

After many years of experimental quest, exciting neutrino 
data pointing towards neutrino oscillations, and thus neutrino masses 
and mixing angles, are starting to be available:  

\begin{itemize}
\item The SuperKamiokande \cite{Superka} data on atmospheric neutrinos 
are interpreted as oscillations of muon neutrinos into
neutrinos that are not $\nu_e$s. Roughly speaking, the measured mixing
angle is close to maximal, $\sin^2 2 \theta > 0.8$, and $\Delta m^2$ is 
in the range $2$ to $6 \times 10^{-3}$ eV$^2$, all at 90\% confidence level. 
\item The solar neutrino deficit is interpreted either as MSW (matter
enhanced) oscillations \cite{MSW} or as vacuum oscillations \cite{osc} 
that deplete the original $ \nu_e$s, presumably in favour of 
$ \nu_\mu$s or alternatively into sterile neutrinos.
The corresponding squared mass differences ($10^{-5}$ to $10^{-4}$ eV$^2$ 
or some $10^{-10}$ eV$^2$) are significantly below the range deduced 
from atmospheric observations. 
\item The LSND signal \cite{LSND} could indicate, 
if confirmed, $\nu_e$s oscillating 
into $ \nu_\mu$s (and $\bar \nu_e$s into $\bar \nu_\mu$s) with 
$\Delta m^2 \sim 0.3-1 eV^2$ and $10^{-3} < \sin^2 2 \theta < 1$ mixing 
angle.
\end{itemize}

The atmospheric plus solar data can be easily accommodated in a three 
family mixing scenario, with two distinct mass differences.
LEP data restricts the number of active neutrino species to three.
Thus, when the LSND signal is also taken into account,
its associated new mass difference requires the
consideration of a supplementary sterile neutrino \cite{sterile}. In 
either case, the physical parameters present in the neutrino Yukawa
sector can be parametrized ``a la CKM'', as in the quark sector, 
with the addition of extra phases if the neutrinos are Majorana particles.

Oscillation experiments are sensitive to mass differences, mixing angles 
and Dirac phases, and not to Majorana phases.
After further confirmation of the present data, a future experimental 
neutrino program should aim at the precise determination of these parameters. 
A {\it neutrino factory} from muon storage rings 
\cite{muring} would provide very intense, pure and flavour-rich 
neutrino beams, well suited for precision studies and even the discovery of 
leptonic CP violation, since the two polarities of the beam are available. 
From a $\mu^-$ beam, the following transitions can be explored:
\begin{eqnarray}
 \mu^- \rightarrow e^-\,  & \nu_\mu &  \, \bar{\nu}_e\, ; \nn\\
& \; & \bar{\nu}_e  \rightarrow \bar{\nu}_e \rightarrow e^+ \;\; 
{\rm disappearance,} \nn \\
& \; & \bar{\nu}_e  \raw \bar{\nu}_\mu \raw \mu^+ \;\; {\rm appearance,}\nn \\
& \; & \bar{\nu}_e  \raw \bar{\nu}_\tau \raw \tau^+ \;\; {\rm appearance}
\;\;\; (\tau^+ \raw \mu^+;\; e^+)\, , \nn \\
&  \nu_\mu  & \raw \nu_\mu\raw \mu^- \;\;\;\;\; {\rm disappearance,}
\nn\\
& \nu_\mu & \raw \nu_e \raw e^- \;\;\;\;\; {\rm appearance,}
\nn\\
& \nu_\mu & \raw \nu_\tau \raw \tau^- \;\;\;\;\; {\rm appearance}
\;\;\; (\tau^- \raw \mu^-;\, e^-)\, .
\label{charges}
\end{eqnarray}
The ``wrong sign'' channels of $\mu^+$, $\tau^+$ and $e^-$ appearance,
for which there would be no beam-induced background at the neutrino factory,
are the good news with respect to other type of neutrino beams.

In this work, we consider convenient parametrizations of the physical 
mixing angles and CP phases in the cases of three and four neutrino species, 
and study their experimental signals at the neutrino factory.

For the sake of illustration, we shall consider as a ``reference 
set-up'' the neutrino beams resulting from the decay of $n_\mu = 2 \times
10^{20} \mu^+$s and/or $\mu^-$s in a straight section of an $E_\mu = 
10\,-\,50$ GeV muon accumulator ring. Muon energies of $40\,-\,50$ GeV are at 
present under discussion \cite{lyon} as a convenient goal, as they allow 
good background rejection \cite{jj} and do not preclude the 
exploration of neutrino signals at lower energies. This is because
the number of neutrinos in a given energy bin does not depend on the 
parent muon energy, while the total number of oscillated and interacted
neutrinos increases with $E_\mu$ \cite{muring}.   

For just three active neutrino species, where the dominant signals peak at
``atmospheric'' distances, we consider a long baseline (LBL) experiment 
located some 732 km downstream, roughly the distance 
from CERN to Gran Sasso or from Fermilab to the Soudan Lab. 
For this scenario we first update the study of CP-odd observables 
performed in \cite{dgh}.
Higher intensity fluxes are also considered for CP-odd signals.
We further discuss the scaling laws that relate the sensitivities 
at different energies and distances.

In the case of three active plus one sterile neutrino, most of the parameter
space can be explored in experiments at short baseline distances (SBL) of
${\cal O}$(1-10 km).

The paper is organized as follows.
In section 2 we update the analysis of the three family scenario and 
derive the scaling laws for the CP-odd observables. 
Section 3 deals with the four species scenario. One and 
two mass scale dominance schemes are derived, and the sensitivity to 
angles and CP-odd phases, relevant in this scheme, are explored. 
We conclude in section 4. 

%
\section{Update of three-family analysis}
%

Let us adopt, from solar and atmospheric experiments,
the indication that $|\Delta m_{solar}^2| \ll |\Delta m_{atm}^2|$, 
dubbed ``minimal '' or ``one mass scale dominance'' scheme. 
Atmospheric or terrestrial experiments have an energy range such that 
$\Delta m^2 L/E_\nu \ll 1$ for the smaller  but not 
necessarily for the larger  of these mass gaps. Even then, 
solar and atmospheric (or terrestrial) experiments are not 
two separate two-generation mixing effects in general. 
Given this hierarchy of mass differences and assigning 
$\Delta m_{solar}^2=\Delta m_{12}^2$, $\Delta m_{atm}^2=\Delta m_{23}^2$,
it is convenient to use the standard 3-family mixing parametrization 
of the PDG, 
\be
U = U_{23} (\theta_{23}) U_{13} (\theta_{13}, \delta ) 
     U_{12} (\theta_{12}) , 
    \label{standpar}
\ee
for the mixing angles and phase. In the approximation $\Delta m^2_{12} = 0$, 
one physical angle and one phase should become unphysical. This is cleanly 
achieved in this parametrization, where the rotation matrix, $U_{12}$, 
associated with the two degenerate eigenstates, is located to the right of 
eq. (\ref{standpar}). In this way, the angle $\theta_{12}$ drops out 
of the neutrino mass matrix when the solar mass difference is
neglected\footnote{Notice that this would not happen automatically
for a different ordering in eq. (\ref{standpar})} and the relevant 
parameters for the atmospheric oscillation are reduced to the smaller set 
$(\theta_{23}, \Delta m^2_{23}$ and $\theta_{13})$. 

From now on we separate the CP-even terms from the CP-odd ones 
in the transition probabilities in the following way, 
\be
P(\nu_\alpha \to \nu_\beta) = \PCPC(\nu_\alpha \to \nu_\beta) +
                              \PCPV(\nu_\alpha \to \nu_\beta).
\label{dos}
\ee

For atmospheric distances, the CP-even components of the transition 
probabilities are accurately given by
\begin{eqnarray}
\PCPC(\nu_e\rightarrow\nu_\mu)&=&  \sin^2(\theta_{23})\, 
\sin^2(2\theta_{13})\,\sin^2\left({\Delta m^2_{23}\, L}\over{4 E_\nu}\right) 
\cr
\PCPC(\nu_e\rightarrow\nu_\tau)&=&   \cos^2(\theta_{23})\, 
\sin^2(2\theta_{13})\,\sin^2\left({\Delta m^2_{23}\, L}\over{4 E_\nu}\right)
\cr
\PCPC(\nu_\mu\rightarrow\nu_\tau)&=&  \cos^4(\theta_{13})\, 
\sin^2(2\theta_{23})\,\sin^2\left({\Delta m^2_{23}\, L}
\over{4 E_\nu}\right)\; ,
\label{cpeven3fam}
\end{eqnarray}
provided the solar mass difference is much smaller than the atmospheric one. 
Generically all flavour transitions occur with the same sinusoidal dependence 
on $\Delta m^2_{23} \, L / E_\nu$. Notice that for $\theta_{13} =0$ the 
probability $P(\nu_\mu \raw \nu_\tau)$ reduces to the two-family (2-3) 
mixing expression, while the other probabilities vanish. 

An analysis of the sensitivity of the neutrino factory to
the transitions above was already done in \cite{dgh}. 
After that work, detailed background estimations have been 
performed \cite{jj}. The overall conclusion 
is that the neutrino factory can reach much higher precision 
in the determination of the parameters involved in the atmospheric 
oscillation than any other 
planned facility. For instance, while all other 
planned experiments will reach at most sensitivities of $\sin^2(\theta_{13}) 
> 10^{-2}$, much lower values are attainable 
at the neutrino factory, down to $\sin^2(\theta_{13}) > 10^{-4}$. This is 
important if the value of this angle turns out to be as small as suggested
by present data, which set the best fit value at 
$\sin^2(\theta_{13}) = 2 \times 10^{-2}$ \cite{fogli}.  

\subsection{CP violation}

The CP-odd terms in eq.~(\ref{dos}) vanish when $\Delta m^2_{12} \, L/E_\nu$ 
effects are neglected, as they should, because the CP-phase of the mixing 
matrix can then be rotated away. The first non-trivial order gives, 
to leading order in the solar mass difference,
\bea
\PCPV(\nu_e \to \nu_\mu) & = & \PCPV(\nu_\mu \to \nu_\tau) 
                         = - \PCPV(\nu_e \to \nu_\tau) \nn \\ 
   & = & - 8 c_{12} c_{13}^2 c_{23} s_{12} s_{13} s_{23} \ \sin \delta
       \left( \frac{\Delta m^2_{12} \, L}{4 E_\nu} \right) \ 
       \sin^2 \left( \frac{\Delta m^2_{23} \, L}{4 E_\nu} \right). 
\label{cpodd3}
\eea
As found in \cite{dgh}, any hope of observability requires that nature chooses 
$\Delta m^2_{12}$ in the higher range allowed by solar experiments, 
$\Delta m^2_{12} \sim 10^{-4} {\rm eV}^2$, and that the ``solar'' mixing 
$\sin (2 \theta_{12})$ is large. These values correspond to the large 
mixing angle solution (LMA-MSW) of the solar deficit. Other analyses of 
the observability of CP violation in the leptonic sector have been presented 
in \cite{romanino} and \cite{barger}.
Previous theoretical work can be found also in \cite{Arafcp}.

We first summarize the results for the CP-odd asymmetry \cite{Nicola} 
obtained in~\cite{dgh}, and update the analysis for different 
$E_\mu$ and higher intensities of the muon beam.

Consider the CP-asymmetry
\begin{equation}
A_{\alpha \beta}^{CP} \equiv \frac{
P(\nu_\alpha\raw \nu_\beta)-P(\bar{\nu}_\alpha \raw \bar{\nu}_\beta)}
{P(\nu_\alpha \raw \nu_\beta)+P(\bar{\nu}_\alpha \raw \bar{\nu}_\beta)} \ ,
\label{CPodd}
\end{equation}
which, in vacuum, would be a CP-odd observable.  Given the present 
experimental constraints, the largest asymmetry is expected in the 
$(e\mu)$-channel. The voyage through our CP-uneven 
planet induces a non-zero $A_{e \mu}^{CP}$ even if CP is conserved, 
since $\nu_e$ and $\bar\nu_e$ are differently affected by the ambient 
electrons \cite{Arafcp}.
In a neutrino factory $A_{e \mu}^{CP}$ would be measured by first
extracting $P(\nu_\mu\raw \nu_e)$ from the produced 
(wrong-sign) $\mu^-$s in a beam from $\mu^+$ decay and 
$P(\bar\nu_e\raw \bar\nu_\mu)$ from the charge conjugate
beam and process. Notice that even if the fluxes are very well
known, this requires a good knowledge of the cross section
ratio $\sigma(\bar\nu_\mu\to\mu^+)/\sigma(\nu_\mu\to\mu^-)$, which
may be gathered in a short baseline experiment. To obtain information on the 
genuinely CP-odd phase, the matter effects in the oscillation probabilities
must also be known with sufficient precision.

A central question on the observability of CP-violation is that of statistics.
In practice, for our reference set-up, there would be too few events to
exploit the explicit $E_\nu$ dependence of the CP-odd effect. 
To construct a realistic CP-odd observable consider the neutrino-energy 
integrated quantity: 
\begin{equation}
{\bar A}^{CP}_{e\mu} (\delta) = \frac{\{{N[\mu^-]}/{N_o[e^-]}\}_{+} 
-\{N[\mu^+]/N_o[e^+]\}_{-}}{\{N[\mu^-]/N_o[e^-]\}_{+} 
+\{N[\mu^+]/N_o[e^+]\}_{-}}\; ,
\label{intasy}
\end{equation}
where the sign of the decaying muons is indicated by a subindex,
$N[\mu^+]$ $(N[\mu^-])$ are the measured number of wrong-sign muons, and 
$N_o[e^+]$ $(N_o[e^-])$ are the expected number of $\bar{\nu}_e (\nu_e)$ 
charged current interactions in the absence of oscillations.
In order to quantify the significance of the signal, we compare the 
value of the integrated asymmetry with its error, in which we include the 
statistical error and a conservative background estimate at
the level of $10^{-5}$.

The measurement of ${\bar A}^{CP}_{e\mu} (\delta)$ at the neutrino 
factory was considered in \cite{dgh} with matter effects taken into account.
It was concluded that, for $2 \times 10^{20}$ useful muons,
if the neutrino mass differences are those indicated by
the ensemble of solar and atmospheric observations and the physics is 
that of three standard families, there was little hope
to observe CP-violation with the beams and detectors under discussion.
Nevertheless, it has been recently pointed out \cite{lyon}  
that it is technically feasible to rise the muon flux intensity by about 
one order of magnitude with respect to our ``reference set-up''.
If the solar LMA solution is confirmed in the coming years, such 
upgrading may allow to unravel CP-violation in the neutrino sector, 
would the CP-odd phase, $\delta$, be sizable.   

\begin{figure}[t]
\begin{tabular}{cc}
\epsfig{file=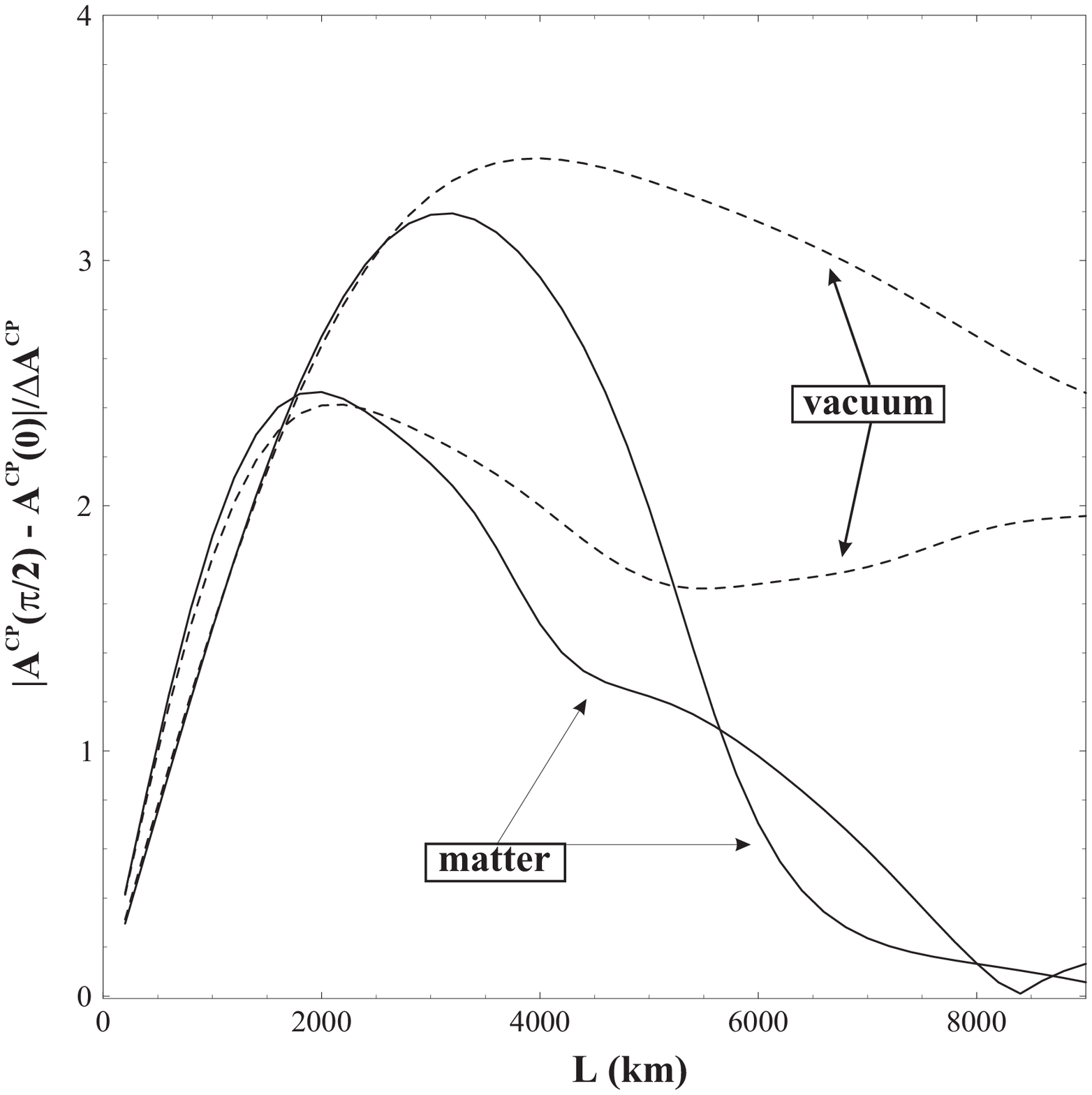,width=7.4cm} & 
\epsfig{file=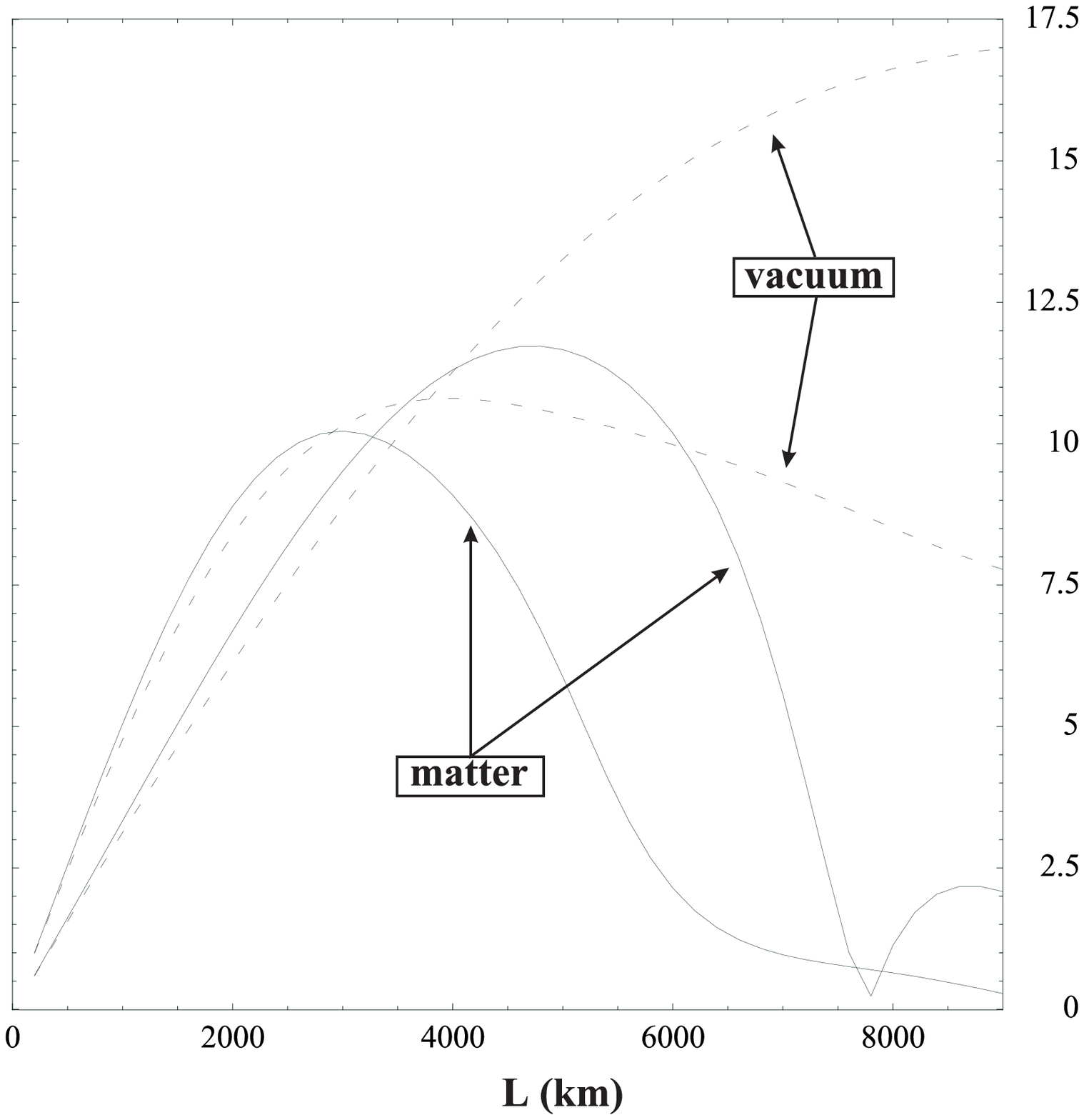,width=7.2cm} \\
\end{tabular}
\caption{\it Signal over statistical uncertainty for $\left| 
{\bar A^{CP}_{e\mu}}(\pi/2) - {\bar A^{CP}_{e\mu}}(0) \right|$ 
as a function of distance. 
Continuous (dashed) lines correspond to matter (vacuum) oscillations. 
In the left side, lower and upper curves correspond to $E_{\mu}=10,~20$ GeV 
for $2 \times 10^{20}$ useful muons/year. In the right the same 
is depicted for $E_{\mu}=20,~50$ GeV and $2 \times 10^{21}$ useful  
muons/year. The chosen CKM parameters are as described in the text.} 
\label{fig:CP}
\end{figure}

The asymmetry in eq. (\ref{intasy})
does not vanish when the CP-phase does due to matter effects.
In order to illustrate the true sensitivity to the CP-odd phase, 
we consider the subtracted asymmetry 
$\left| {\bar A^{CP}_{e\mu}}(\pi/2) - {\bar A^{CP}_{e\mu}}(0) \right|$, 
i.e. the difference between the asymmetry at the maximum value of the CP-phase, 
$\delta=\pi/2$, and that at $\delta = 0$.  Of course, if the 
subtracted term is large, performing this theoretical subtraction 
in practice would require a very precise knowledge of matter effects 
and also of the remaining oscillation parameters. Fig.~\ref{fig:CP} displays 
the signal-over-noise ratio for the subtracted asymmetry as 
a function of distance, for $2 \times 10^{20}$ (left) and $2 \times 10^{21}$ 
(right) useful muons and for a $10$ kTon detector. 
The oscillation parameters in Fig.~\ref{fig:CP} are chosen to be:  
$\Delta m^2_{12} = 10^{-4}$ eV$^2$, $\Delta m^2_{23} = 2.8 \times 10^{-3}$ 
eV$^2$, $\theta_{12} = 22.5^\circ$, $\theta_{13} = 13^\circ$ and $\theta_{23} = 
45^\circ$ with the phase $\delta = -\pi/2$ in the convention of the PDG.  
The value chosen for $\Delta m^2_{23}$ is the central one of the most recent 
SuperKamiokande analysis~\cite{Superka} \cite{fogli}: $2.8 \times 10^{-3}$ 
eV$^2$. When obtaining this figure (as well as in the following ones 
concerning CP-violation) the exact theoretical expressions have been used.  

For $2 \times 10^{21}$ useful muons, the number of ``standard deviations'' 
is seen to exceed a comfortable $\sim 5$ at a large range of distances, 
with a peak towards $3000$ km. 

Let us discuss the relative size of the matter induced asymmetry versus 
the true CP signal (i.e. the terms in the asymmetry proportional to 
$\sin \delta$). For $\theta_{13}=13^\circ$ (which is the present upper
bound from Chooz \cite{chooz}) and at moderate energies, $E_\mu=10, 20$ GeV, 
its contribution is $3-4$ times larger than the true CP-signal, which is 
around $5\%-10\%$, and thus it is necessary to know the subtracted term with 
a precision better than $20\%$ in order to extract any information on $\delta$.
At higher energy, $E_{\mu}=50$ GeV, the situation gets worse: the matter 
induced asymmetry turns out to be much larger than the true 
CP-signal, which is of the order of a few per cent. It does not seem realistic 
to expect to know the error in the (by then hypothetically measured) leptonic 
parameters with sufficient precision to measure the true CP signal in this case. 
On the other hand, analyzing the energy dependence of the signal may be 
feasible at the larger statistics available at this high energy, which might 
allow a cleaner separation of matter effects from CP violating ones. This 
possibility will be studied in the future.

It is important to realize, however, that the relative size of the true CP 
signal and the total asymmetry depends very much on the value of the angle 
$\theta_{13}$ in vacuum\footnote{A more detailed discussion of the dependence 
on $\theta_{13}$ and matter effects will be presented elsewhere \cite{ourprocs}.}. 
As $\theta_{13}$ is decreased, not only the term proportional to $\sin \delta$ 
in the asymmetry increases\footnote{Notice however, that the ratio of the 
asymmetry to its statistical error does not change, so $2 \times 10^{21}$ useful 
muons would still be needed.} \cite{romanino}, but also the matter induced 
asymmetry decreases. As an example, for $E_\mu=20$ GeV and $\theta_{13}=5^\circ$, 
the matter induced asymmetry is of the same order as the true CP-signal 
and becomes five times smaller at $\theta_{13}=1^\circ$. At this
value, the true CP asymmetry becomes as large as $50\%$. The subtraction 
of the small matter induced asymmetry in this situation should be very easy. 
One should keep in mind, that although the present best fit value for 
$\theta_{13}$ is $8^\circ$, a much smaller value would also be in perfect 
agreement with atmospheric and solar data.

\begin{figure}[t]
\begin{center}
\epsfig{file=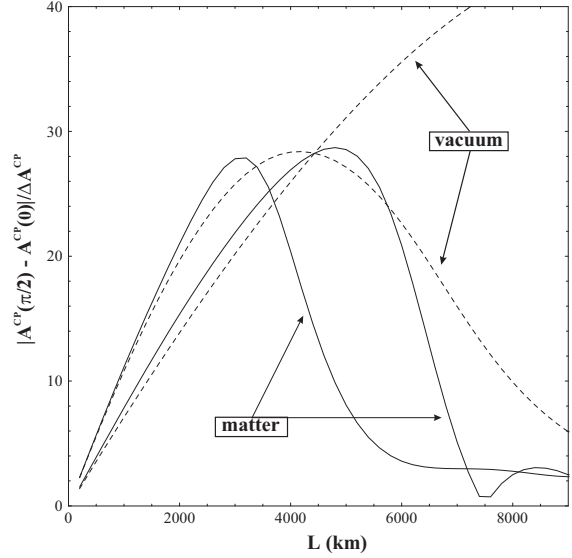,width=7.4cm} 
\end{center}
\caption
{\it Signal over statistical uncertainty for $\left| 
{\bar A^{CP}_{e\mu}}(\pi/2) - {\bar A^{CP}_{e\mu}}(0) \right|$ as a function 
of distance, for $\Delta m^2_{12} = 8 \times 10^{-4}$ eV$^2$ and the rest of 
the parameters as in the left plot of Fig. 1.}
\label{fig:CPclor}
\end{figure}

Up to here we have considered the experimental constraints resulting 
from the ensemble of the solar data. The CP-violation effects are much 
bigger for the larger mass differences that become possible if
the results of some solar neutrino experiment are disregarded. This is 
because the CP asymmetry grows linearly with the solar mass difference. 
As an example, increasing $\Delta m^2_{12}$ from $10^{-4}$ eV$^2$ to 
$8 \times 10^{-4}$ eV$^2$, with the other parameters fixed as in the left 
plot of Fig. 1,  results in a very promising signal over noise ratio even
in the ``conservative'' option of $2 \times 10^{20}$ useful muons.
This is depicted in Fig. \ref{fig:CPclor}.

In ref. \cite{dgh} the T-odd asymmetry at the neutrino factory was as well 
briefly discussed. As it requires the measurement of the electron charge, 
which seems experimentally out of reach at present, we refrain from 
its further study in this paper.

\subsection{Matter effects and scaling laws for CP-observables}
%
It is interesting to understand how the sensitivities to mixing angles and 
CP-odd effects scale with LBL distances and beam energies. We concentrate 
here on CP-odd observables, since the CP-even ones were discussed in
\cite{dgh}.

Consider the vacuum CP-odd observable defined in eq.~(\ref{intasy}). 
Neglecting the detection background and other systematic errors, the 
signal over its statistical error, depicted in Fig.~\ref{fig:CP}, scales as
\be
\frac{A_{e \mu}^{CP}}{\Delta A_{e \mu}^{CP}} = 
\frac{ \PCPV \sqrt{N_{CC}} }{\sqrt{\PCPC}} \propto \sqrt{E_\nu}  
\left | \sin \left ( \frac{\Delta m^2_{23} \, L}{4 E_\nu} \right ) \right |,
\label{scaling}
\ee
where we have used that the number of charged currents scales as $N_{CC} 
\propto E_\nu^3/L^2$. Hence, the location of the maximum is at 
$\Delta m^2_{23} \, L/E_\nu = 2 \pi$ whereas the height scales with the 
square root of the energy. 

In matter, the situation is somewhat more involved. 
First, the splitting between mass eigenstates in vacuum must be replaced by
the splitting of the mass eigenstates in matter, thus affecting the
location of the maximum in the signal-to-noise ratio. 
The second and most important difference is that the effective 
mixing angles in matter depend on the energy, thus modifying the 
scaling with the energy of the maximum height. In the regime in which 
the matter-induced mass splitting, $A = 2 E_\nu \sqrt{2} G_F n_e$ (where 
$n_e$ is the number density of electrons in the earth) is larger than 
all mass differences ($\Delta m^2_{12}$ and $\Delta m^2_{23}$) it is easy to 
show that only the effective angle $\theta_{13}$ changes significantly with 
the energy. More specifically, for the signal-to-noise ratio the angular 
difference between vacuum and matter relies on the effective $\cos 
(\theta_{13})$, which has a $1/\sqrt{E_\nu}$ functional dependence on energy. 
This factor cancels the $\sqrt{E_\nu}$ dependence of the signal-to-noise 
ratio in vacuum, and the height of the maximum does no longer scale with 
the energy. This behaviour can be seen in Fig. \ref{fig:CP} where the 
signal-to-noise ratio is plotted both in vacuum and in matter at different 
values of the energy of the muon beam. 

Table \ref{tab:scaling} summarizes the scaling laws of height and location 
of the CP-violating signal-to-noise ratio maximum as a function of $E_\nu$ 
and $L$, both in vacuum and in matter.
%
%
\begin{table}[t]
\centering
\begin{tabular}{||c|c|c||}
\hline\hline
         & Height & Location \\
\hline \hline & & \\
Vacuum  & $\propto \sqrt{E_\nu}$ & $\dd L = \frac{2 \pi E_\nu}{\Delta m^2_{23}}$ \\
& & \\ \hline & & \\
Matter & $\dd \propto 1 $ & $L = \dd \frac{\pi}{A}$ \\
& & \\ \hline \hline
\end{tabular}
\caption{\it{
Height and location of the maximum of the CP-violating signal-to-noise ratio 
as a function of the neutrino energy $E_\nu$ and of the source-to-detector 
distance $L$ in vacuum, and in matter for $A \gg \Delta m^2_{ij}$.}}
\label{tab:scaling}
\end{table}
%
%
%
\section{Four-Neutrino species}
\label{sect:fourneutrino}
%
%
The LSND signal of $\nu_\mu \to \nu_e$ oscillations would indicate 
a third distinct neutrino mass range, $\Delta m_{LSND}^2 = 0.3 - 1$ 
eV$^2$. It is then necessary to introduce a fourth light neutrino, a 
sterile one, in order to comply with the bounds on the number of neutrinos 
coupled to the $Z^0$ \cite{LEPnu}.
\begin{figure}[t]
\begin{tabular}{lcr}
\epsfig{file=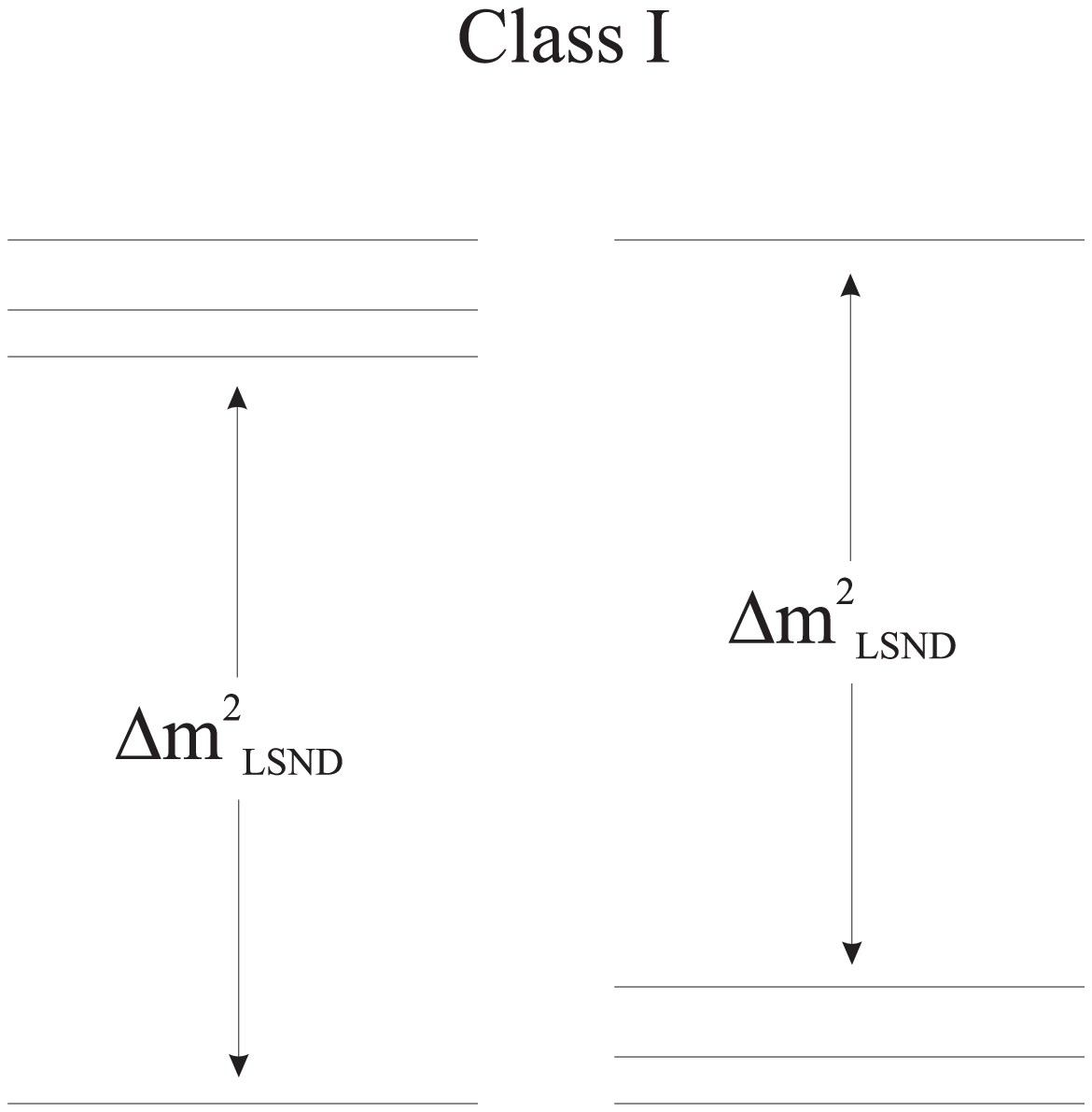, width=6.5cm} & \mbox{ \hskip 1cm} &
\epsfig{file=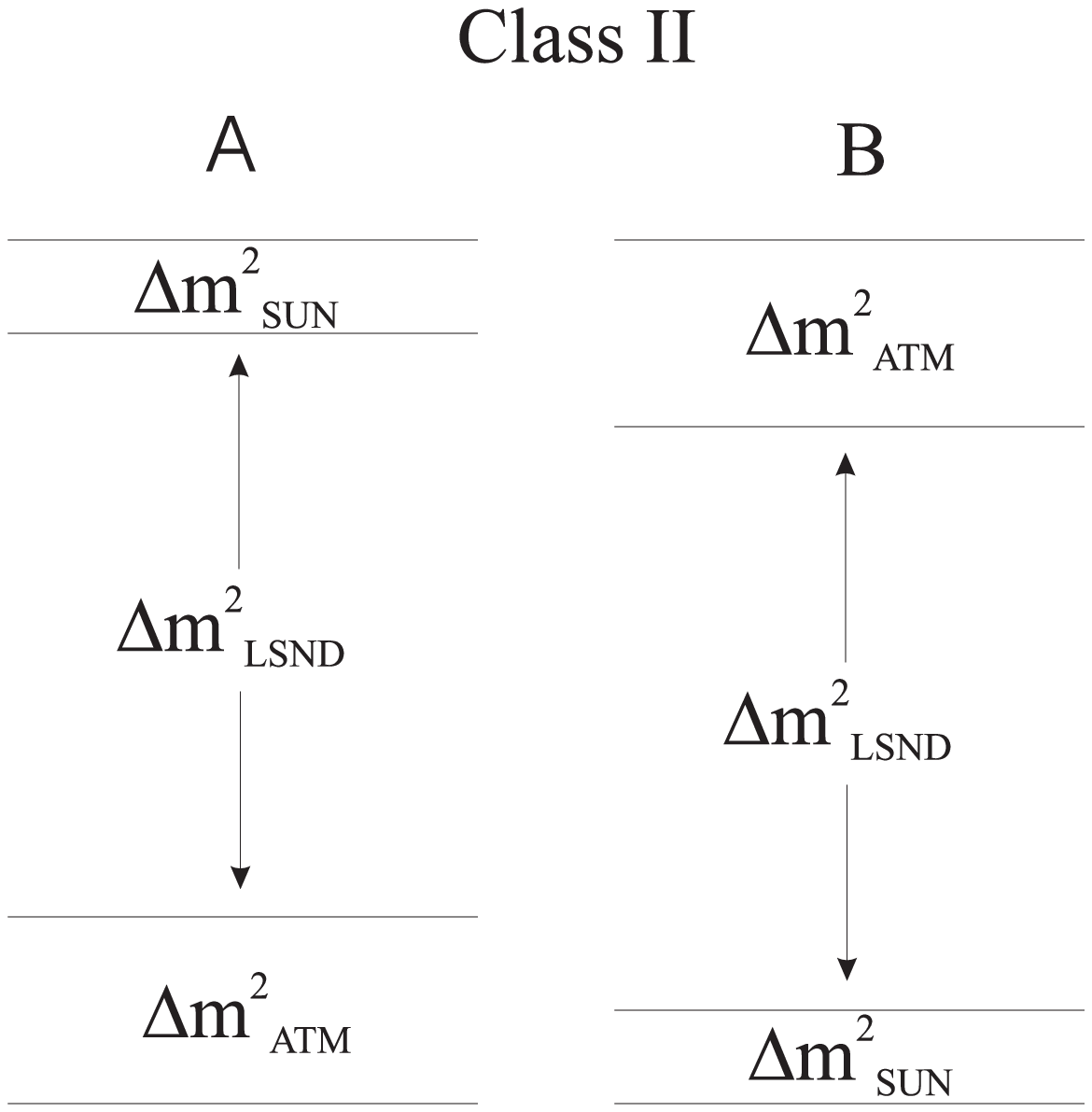, width=6.5cm} 
\end{tabular}
\caption{{\it Different type of four-neutrino families scenarios: 
Class-I scenarios (left); Class-II scenarios (right).}} 
\label{fig:4famtyp}
\end{figure}

A combined analysis of the results for solar, atmospheric and LSND 
experiments points \cite{giunti} towards a four neutrino pattern in which 
there are two nearly degenerate neutrino pairs separated by a large mass gap, 
$\Delta m^2_{LSND}$. Different scenarios are depicted in Fig.~\ref{fig:4famtyp}. 
Our analysis sticks to class-II, since class-I seems to be excluded by 
data~\cite{bilenky}.

Whatever the mechanism responsible for the neutrino masses,
given $n$ light neutrino species, oscillation experiments are 
only sensitive to a unitary $n \times n$ mixing matrix ``a la CKM'': 
the effects of Majorana phases being suppressed by factors of 
$m_\nu/E_\nu$. In all generality, the parameter space of a 
four-species scenario would consist of six rotation angles and three 
complex phases if the neutrinos 
are Dirac fermions, while it spans six angles plus six phases if the 
neutrinos are Majorana fermions. Among the latter, three are pure 
Majorana phases and thus to be disregarded in what follows, reducing the 
analysis to the mentioned $4 \times 4$ ``Dirac-type'' system. 
 
We assign the mass eigenstates in the following way:
\be
\Delta m_{sol}^2 = \Delta m_{12}^2 \ll 
\Delta m_{atm}^2 = \Delta m_{34}^2 \ll 
\Delta m_{LSND}^2 = \Delta m_{23}^2.    
\label{hier4fam}
\ee 
Given the large hierarchy indicated by data, two approximations are useful:
\begin{enumerate}
\item
$\Delta m_{12}^2 = \Delta m_{34}^2 = 0$, ``one mass scale dominance'' (or 
minimal) scheme,   
\item
$\Delta m_{12}^2 = 0$, ``two mass scale dominance'' (or next-to-minimal) 
scheme.  
\end{enumerate}
The number of independent angles and phases is then reduced as reported in 
Tab.~\ref{tab:4fampar}. The minimal scheme is sufficient to illustrate 
the sensitivity to the mixing angles at the neutrino factory. In this 
approximation, a physical CP-odd phase is still present although not 
sufficient to produce CP-violation effects in neutrino oscillations. 
The next-to-minimal scheme is necessary to address the question of 
CP-violation, as two non-zero mass differences are required for 
producing observable effects, alike to the standard three-family scenario. 

%
%
\begin{table}[t]
\centering
\begin{tabular}{||c|c|c|c||}
\hline\hline
         & Angles & Dirac CP-phases & Majorana CP-phases \\
\hline\hline
& & & \\
Majorana $\nu$'s &   6   &   3   &  3   \\
& & & \\
\hline 
& & & \\
Dirac $\nu$'s &   6   &   3   &  0   \\
& & & \\
\hline
& & & \\
Dirac $\nu$'s   &   5   &   2   &  0   \\
$\Delta m_{12}^2 = 0$ & & & \\
& & & \\
\hline
& & & \\
Dirac $\nu$'s   &   4   &   1   &  0   \\
$\Delta m_{12}^2 = \Delta m_{34}^2 = 0$ & & & \\
& & & \\
\hline \hline
\end{tabular}
\caption{\it{Parameter space for four neutrino families: for Dirac
neutrinos we consider the general case with three non-zero mass differences
and the particular case (considered in the rest of the paper) with
one or two mass differences set to zero; for Majorana neutrinos we
consider only the general case.}}
\label{tab:4fampar}
\end{table}
%
%

A general rotation in a four dimensional space can be obtained by performing
six different rotations $U_{ij}$ in the $(i,j)$ plane, resulting in plenty 
of different possible parametrizations of the mixing matrix, disregarding 
phases. We choose the following convenient parametrization, given the 
hierarchy of mass differences of eq. (\ref{hier4fam}):
\be
U = U_{14} (\theta_{14}) U_{13} (\theta_{13}) U_{24} (\theta_{24}) 
    U_{23} (\theta_{23},\delta_3) U_{34} (\theta_{34}\delta_2) 
    U_{12} (\theta_{12},\delta_1).
\label{ourpar}
\ee
As shown in Table 2, if a given mass difference vanishes the number of
physical angles and phases gets reduced by one. A convenient parametrization 
of the angles is that in which the rotation matrices corresponding to the 
most degenerate pairs of eigenstates are placed to the extreme right. If the 
eigenstates $i$ and $j$ are degenerate and the matrix $U_{ij}$ is located
to the right in eq.~(\ref{ourpar}), the angle $\theta_{ij}$ becomes 
automatically unphysical in this parametrization. If a different ordering is 
taken no angle disappears from the oscillation probabilities. A redefinition of
the rest of the parameters would then be necessary in order to illustrate the 
remaining reduced parameter space in a transparent way.  
Our parametrization corresponds thus to the ``cleanest'' choice, having settled 
at the extreme right the rotations corresponding to the most degenerate pairs. 

In the ``one mass dominance'' scheme,  the pairs $(\theta_{12}, \delta_1)$ and 
$(\theta_{34}, \delta_2)$ decouple automatically. In the ``two mass dominance''  
scheme, only the pair $(\theta_{12}, \delta_1)$ does. Thus only the exact 
number of physical parameters, according to Table 2, remains both in the 
minimal and next-to-minimal schemes. Notice that it is also important to 
distribute the phases so that they decouple, together with the angles, 
when they should.  

\subsection{Sensitivity reach of the neutrino factory for four neutrino species}

We concentrate now on the sensitivity to the different angles describing 
the system when only the big LSND mass difference is taken into account, that
is  the ``one mass scale'' approximation, discussed at the beginning of 
this section.
Four rotation angles ($\theta_{13}$, $\theta_{14}$, $\theta_{23}$ and 
$\theta_{24}$) and one complex phase ($\delta_3$) remain. The two rotation 
angles that have become unphysical are already tested at solar ($\theta_{12}$ 
in our parametrization) and atmospheric ($\theta_{34}$) neutrino experiments. 
The remaining four can be studied at the neutrino factory with high precision, 
due to the rich flavour content of the neutrino beam. Notice that the number 
of flavour transitions that can be measured is enough to constraint all these 
angles. We will concentrate for illustration on the following channels:
\bea
  \bar{\nu}_e & \to  \bar{\nu}_\mu & \to  \mu^+
                             \qquad (\mu^+ {\rm appearance}) \nn \\
  \nu_\mu & \to  \nu_\mu & \to  \mu^- 
                             \qquad (\mu^- {\rm disappearance}) \nn \\
  \bar{\nu}_e & \to  \bar{\nu}_\tau & \to  \tau^+ 
                             \qquad (\tau^+ {\rm appearance}) \nn \\
  \nu_\mu & \to  \nu_\tau & \to  \tau^- 
                             \qquad (\tau^- {\rm appearance}). 
\eea
The corresponding probability transitions acquire a simple form:
\bea
\label{cpeven1}
\PCPC(\nu_e \to \nu_\mu) &=& 4 c^2_{13} c^2_{24} c^2_{23} s^2_{23} \
\sin^2 \left ( \frac{\Delta m_{23}^2 L}{4 E} \right )  \ , \\
\PCPC(\nu_\mu \to \nu_\mu) & = & 1
 - 4 c^2_{13} c^2_{23} ( s^2_{23} + s^2_{13} c^2_{23} ) \
       \sin^2 \left ( \frac{\Delta m_{23}^2 L}{4 E} \right ) \ ,
\label{cpeven2} \\
\PCPC(\nu_e \to \nu_\tau) & = & 4 c^2_{23} c^2_{24} 
\left [ ( s^2_{13} s^2_{14} s^2_{23} + c^2_{14} c^2_{23} s^2_{24} ) 
\right .  \nn \\
& - & \left .  2 c_{14} s_{14} c_{23} s_{23} s_{13} s_{24} 
\cos \delta_3 \right ] \
\sin^2 \left ( \frac{\Delta m_{23}^2 L}{4 E} \right )  \ ,
\label{cpeven3} \\
\PCPC(\nu_\mu \to \nu_\tau) & = & 4 c^2_{13} c^2_{23} 
\left [ ( s^2_{13} s^2_{14} c^2_{23} + c^2_{14} s^2_{23} s^2_{24} ) 
\right . \nn \\
& + & \left .  2 c_{14} s_{14} c_{23} s_{23} s_{13} s_{24} 
\cos \delta_3 \right ] \
\sin^2 \left ( \frac{\Delta m_{23}^2 L}{4 E} \right ).
\label{cpeven4}
\eea
Notice that the physical phase appears in $\PCPC(\nu_e \to \nu_\tau)$ and 
$\PCPC(\nu_\mu \to \nu_\tau)$ in a pure cosine dependence. Actually, no 
CP-odd observable can be built out of the oscillation 
probabilities in this approximation in spite of the existence of a
physical CP-odd phase in the mixing matrix. 

The existing experimental data impose some constraints on the parameter 
space, but still leave free a large range of angles and phases. 

Bugey and Chooz are sensitive to the oscillation $\bar{\nu}_e \to \bar{\nu}_e$,
\be
P(\bar{\nu}_e \to \bar{\nu}_e) = 1
 - 4 c^2_{23} c^2_{24} ( s^2_{24} + s^2_{23} c^2_{24} ) 
       \sin^2 \left ( \frac{\Delta m_{23}^2 L}{4 E} \right ).
\ee
The resulting bound is 
\be
( c_{23}^2 \sin^2 2 \theta_{24} + c^4_{24} \sin^2 2 \theta_{23}  ) \le 0.2 \ ,
\label{boundbugey}
\ee
where Bugey gives a slightly stronger constraint in the larger mass range 
allowed by LSND. In our computations we safely stay within both experimental 
constraints. Notice that in the assumption of small angles,  
this bound forces the mixings $s^2_{23}$ and $s^2_{24}$ to be small 
and leaves more freedom in the mixings of the sterile neutrino: 
$s_{13}^2$ and $s_{14}^2$.

On the other hand, the LSND signal of $\nu_e \to \nu_\mu$ transitions 
indicates a bound for the combination $10^{-3} \leq c^2_{13} c^2_{24} 
\sin^2 2 \theta_{23} \leq 10^{-2}$, depending on the LSND mass difference. 
This bound fits nicely with the Chooz constrain to select a small $s_{23}^2$.

We choose to be ``conservative'', or even ``pessimistic'', in order to 
illustrate the potential of the neutrino factory. In the numerical 
computations below we will make the assumption that all angles crossing 
the large LSND gap, $\theta_{13}$, $\theta_{14}$, $\theta_{23}$ and 
$\theta_{24}$ are small.
  
The large LSND mass difference, $\Delta m_{23}^2 \simeq 1 {\rm eV}^2$, 
calls for a SBL experiment rather than a LBL one. For illustration we 
consider in what follows an hypothetical 1 Ton detector located at 
$\simeq 1$ km distance from the neutrino source. We assume that the detector 
has $\tau$ tracking and $\mu$ and $\tau$ charge identification capabilities. 
We consider a muon beam of $E_\mu = 20$ GeV, resulting in $N_{CC} \simeq 10^7$ 
charged leptons detected, for a beam intensity of $2 \times 10^{20}$ useful 
$\mu^-$ per year. An efficiency of $\epsilon_\mu = 0.5, \epsilon_\tau = 0.35$ 
for $\mu$ and $\tau$ detection respectively, and a background contamination 
at the level of $10^{-5} N_{CC}$ events are included. 

\begin{figure}[t]
\vspace{0.1cm}
\centerline{
\epsfig{figure=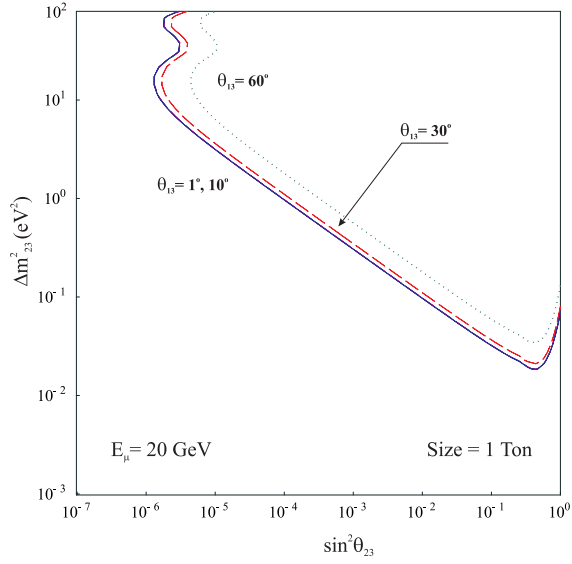,height=7.4cm,angle=0}}
\caption{\it{
Sensitivity reach in the 
$s^2_{23} / \Delta m_{23}^2$ plane at different values of 
$\theta_{13}= 1^\circ, 10^\circ, 30^\circ$ and $60^\circ$
for $\mu^+$ appearance. We consider a 1 Ton detector at $ 1$ km 
from the source and $2 \times 10^{20}$ useful muons/year.}} 
\label{fig:mu23a}
\end{figure}

\begin{itemize}
\item {\bf $\theta_{23}$ and $\theta_{13}$ from $\mu$ channels}
\end{itemize}
The $\mu^+$ appearance channel is particularly sensitive to $\theta_{23}$. 
Fig. \ref{fig:mu23a} shows the sensitivity reach 
in the $s^2_{23} / \Delta m_{23}^2$ plane for different
values of $\theta_{13}$. Inside the LSND allowed region the dependence 
on $\theta_{13}$ is mild: $s_{23}^2$ can reach $10^{-6}$ for $\theta_{13} 
\simeq 1^\circ$ or $6 \times 10^{-6}$ for $\theta_{13} \simeq 60^\circ$. 

\begin{figure}[t]
\vspace{0.1cm}
\begin{tabular}{cc}
\epsfig{figure=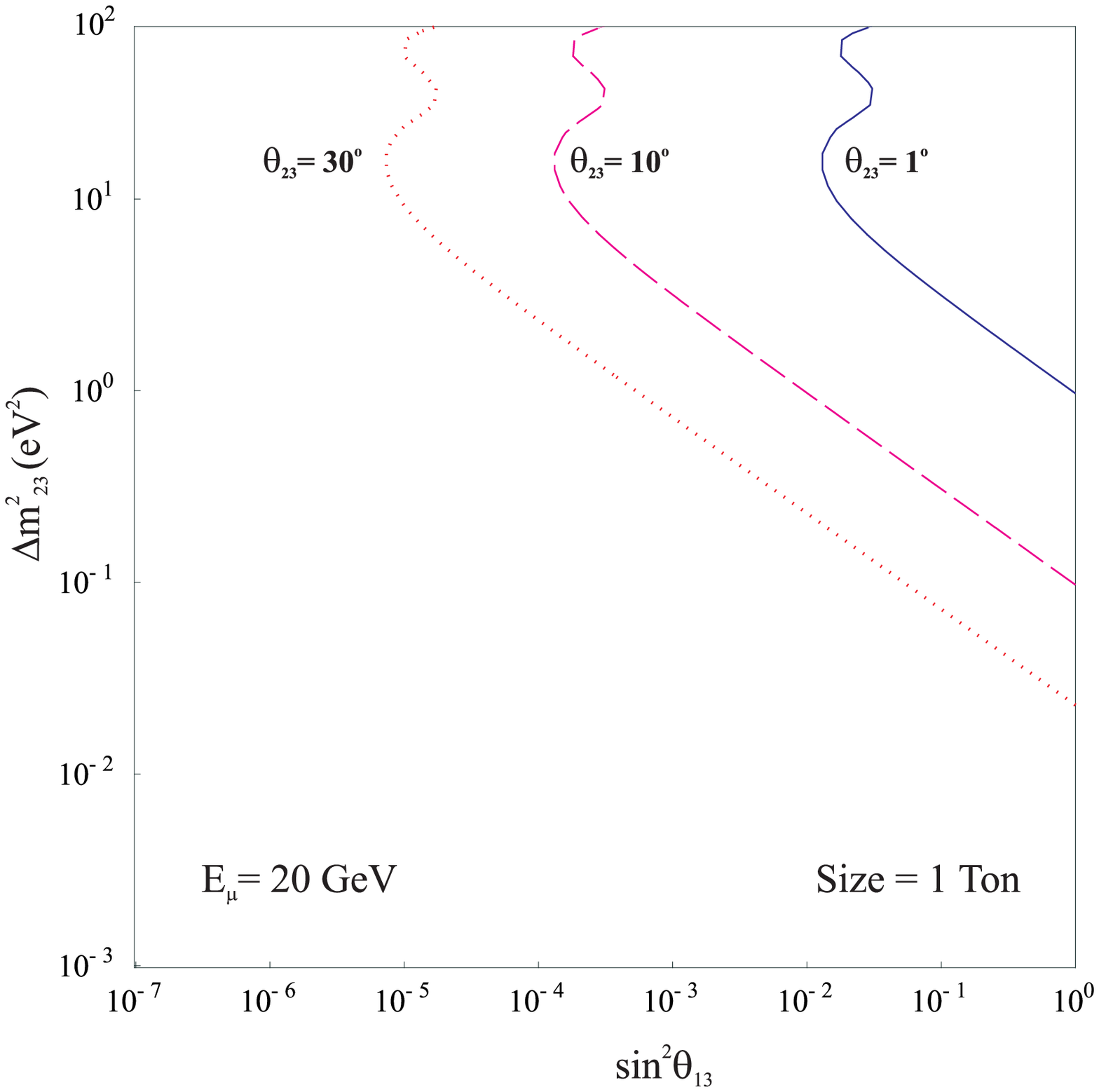,height=7.4cm,angle=0} &
\epsfig{figure=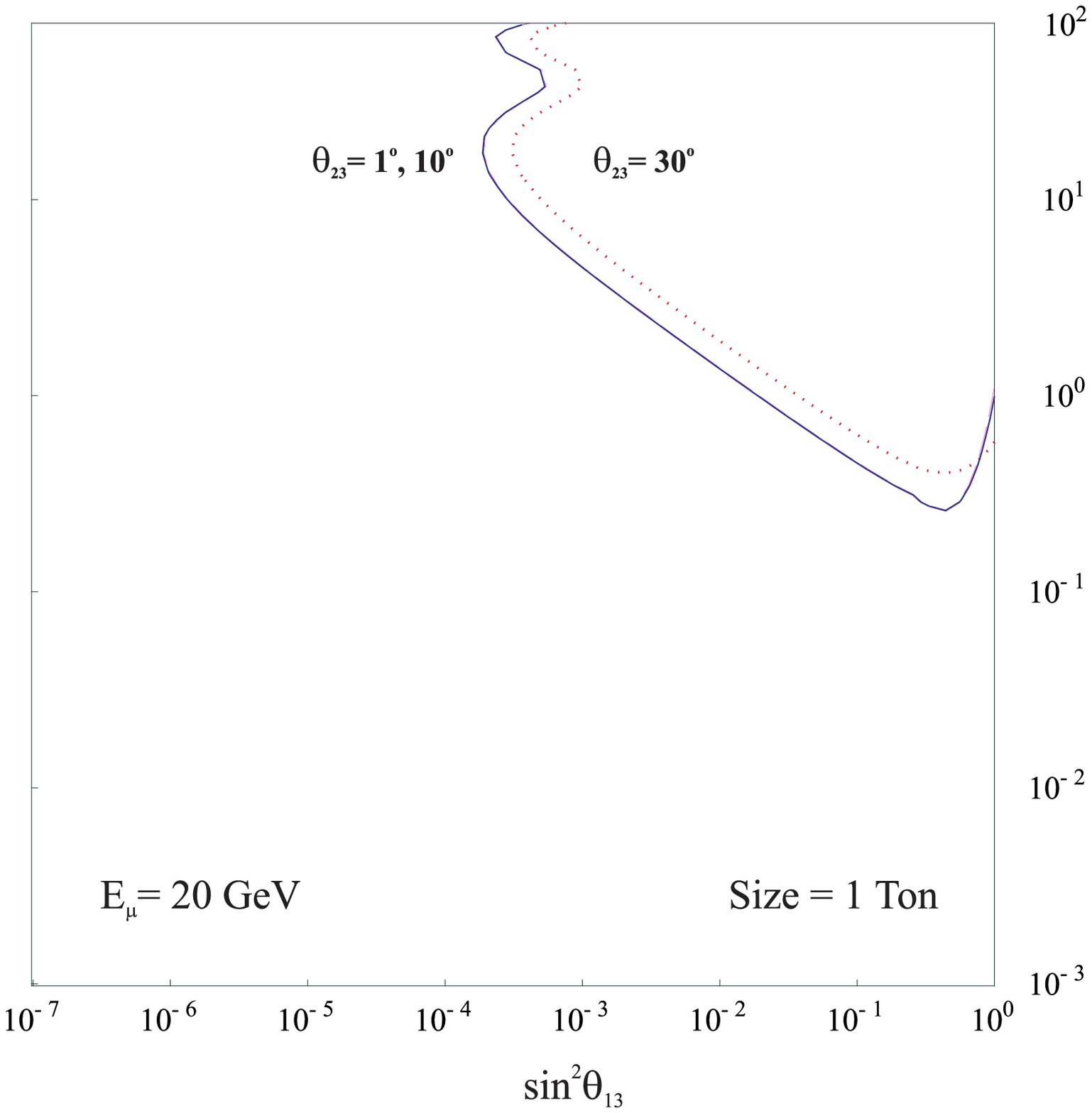,height=7.4cm,angle=0} 
\end{tabular}
\caption{\it{
Sensitivity reach in the 
$s^2_{13} / \Delta m_{23}^2$ plane at different values of 
$\theta_{23}= 1^\circ, 10^\circ$ and $30^\circ$ for $\mu^+$ appearance 
(left) and disappearance (right). We consider a 1 Ton detector at $ 1$ km 
from the source and $2 \times 10^{20}$ useful muons/year.}} 
\label{fig:mu13ad}
\end{figure}

Concerning the sensitivity to $\theta_{13}$, Fig. \ref{fig:mu13ad} (left) 
illustrates the reach from $\mu$ appearance measurements in the 
$s^2_{13} / \Delta m_{23}^2$ plane, for different values of 
$\theta_{23}$. Inside the LSND allowed region, the sensitivity 
to this angle strongly depends on the value
of $\theta_{23}$, with the larger sensitivity attained for large 
values of $\theta_{23}$, a scenario somewhat disfavoured by the LSND 
measurement. For small values of $\theta_{23}$ ($\simeq 1^\circ$), 
the smallest testable value of $s_{13}^2$ is $\sim 10^{-2}$. 
Nevertheless, in this range the muon disappearance channel proves 
quite more sensitive: Fig.~\ref{fig:mu13ad} (right) goes down to 
$s_{13}^2$ as small as $10^{-4}$ for $\theta_{23} \simeq  1^\circ$.

\begin{figure}[t]
\vspace{0.1cm}
\centerline{
\epsfig{figure=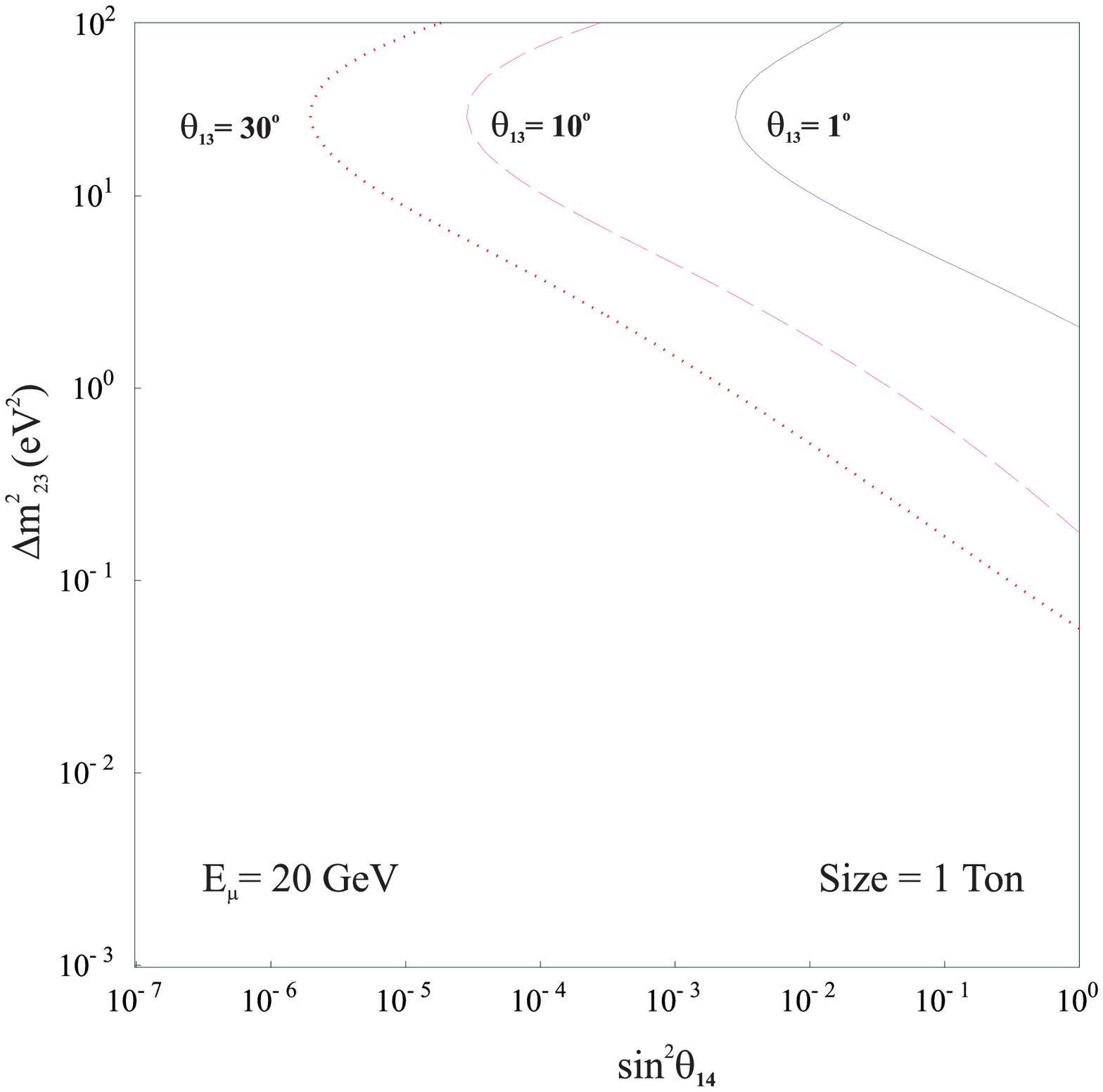,height=7.4cm,angle=0}} 
\caption{\it{
Sensitivity reach in the $s^2_{14}/ \Delta m_{23}^2$ plane at 
different values of $\theta_{13}= 1^\circ, 10^\circ$ and $30^\circ$
for $\tau^-$ appearance. We consider a 1 Ton detector at $1$ km 
from the source and $2 \times 10^{20}$ useful muons/year.}} 
\label{fig:taum14a}
\end{figure}
\begin{figure}[th]
\vspace{0.1cm}
\begin{tabular}{cc}
\epsfig{figure=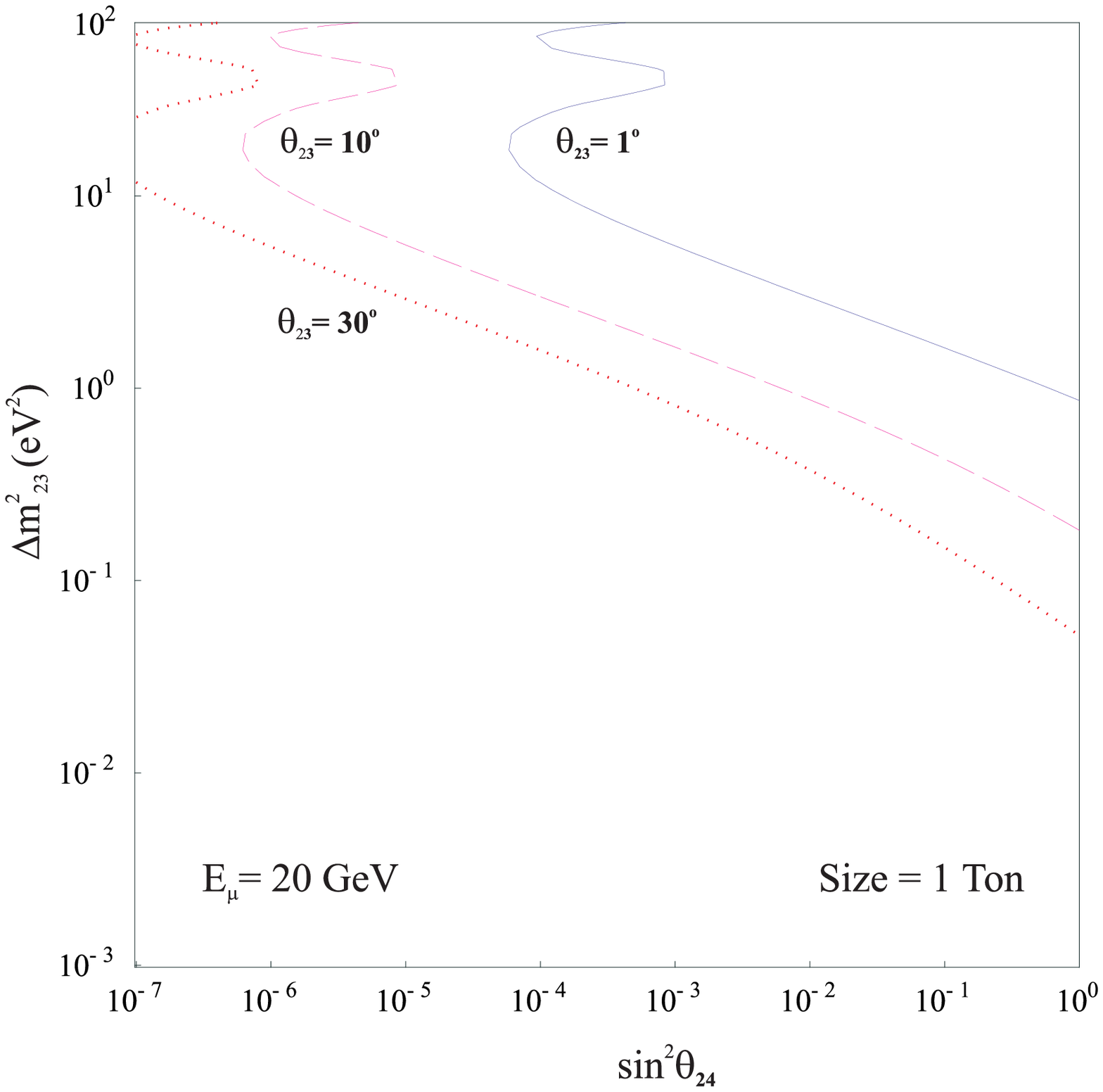,height=7.4cm,angle=0} & 
\epsfig{figure=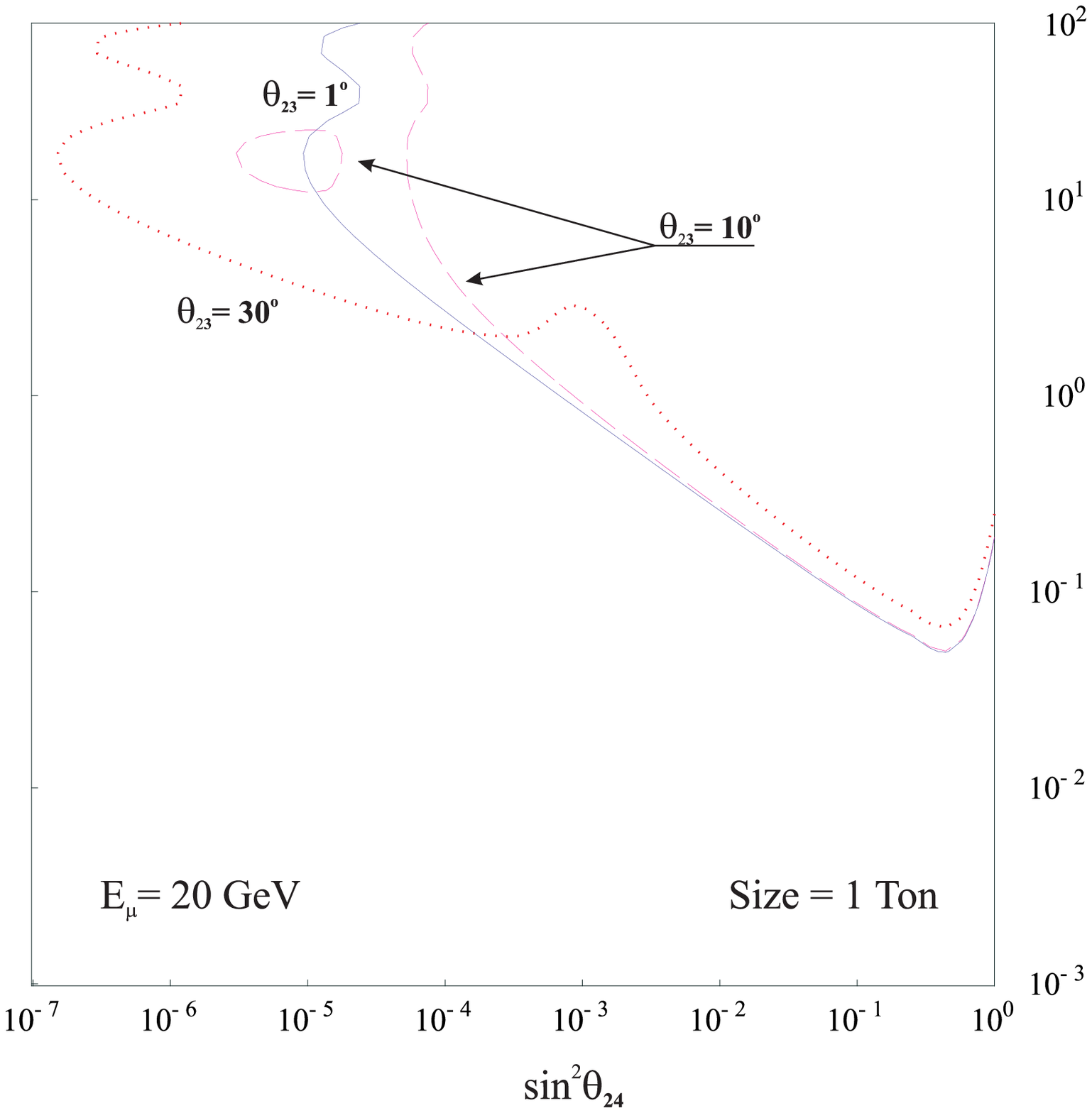,height=7.4cm,angle=0} 
\end{tabular}
\caption{\it{
Sensitivity reach in the $s^2_{24} / \Delta m_{23}^2$ plane at 
different values of $\theta_{23}= 1^\circ, 10^\circ$ and $30^\circ$ 
$\tau^+$ appearance. We consider a 1 Ton detector at $1$ km 
from the source and $2 \times 10^{20}$ useful muons/year.}} 
\label{fig:tauem24a}
\end{figure}

\begin{itemize}
\item  {\bf $\theta_{14}$ and $\theta_{24}$ from $\tau$ channels}
\end{itemize}

The $\tau^-$ appearance channel is quite sensitive to both $s_{14}^2$ and 
$s_{24}^2$. Fig. \ref{fig:taum14a} illustrates the sensitivity to $s_{14}^2$
as a function of $\theta_{13}$: for about $1^\circ$, sensitivities of 
the order of $10^{-2}$ are attainable, while for $10^\circ$ the 
reach extends to $4 \times 10^{-5}$. For even larger values of $\theta_{13}$ 
it goes down to $10^{-6}$ (we recall that $\theta_{13}$ is not constrained by 
the present experimental bounds).

Fig. \ref{fig:tauem24a} (left) depicts the foreseeable sensitivity reach to 
$s^2_{24}$ as a function of $\theta_{23}$ : for small values of 
$\theta_{23}$ the sensitivity to $s_{24}^2$ goes down to $10^{-6}$. 

In contrast, the $\tau^+$ appearance channel looks less promising.
This is illustrated in Fig. \ref{fig:tauem24a} (right): due 
to the relative negative sign between the two terms in the 
analytic expression for $P(\nu_e \to \nu_\tau)$, eq.~(\ref{cpeven3}),
cancellations for particular values of the angles occur, resulting
in a decreasing sensitivity in specific regions of the parameter space.
For instance, for $\theta_{23} = 10^\circ$, the reach in 
$s^2_{24}$ splits into two separate regions for the LSND allowed range 
$\Delta m^2_{23} \sim 10^1$ eV$^2$.
This sensitivity suppression is absent in the $\tau^-$ channel as the
relative sign between the two terms in $P(\nu_\mu \to \nu_\tau)$, 
eq.~(\ref{cpeven4}), is positive.

The overall conclusion of this analysis is that, in the minimal scheme 
for four-neutrino families, 
a 1 Ton near detector with $\mu$ and $\tau$ charge identification 
is suitable to fully explore the CP-even part of the whole parameter space. 

\subsection{CP Violation with four light neutrino species}
\label{sect:4CPviolation}
%
As in the standard three-family scenario, in order to reach observable 
CP-odd effects in oscillations it is necessary to have both physical CP-odd 
phases and at least two non-vanishing mass differences. The next-to minimal 
or ``two mass scale dominance'' scheme, described at the beginning of this
section, is thus suitable.

As explained above, the parameter space consists of five angles and two 
CP-odd phases. Expanding the transition probabilities to leading order in 
$\Delta m^2_{atm}$ (i.e. $\Delta m^2_{34}$ in our parametrization), it follows 
that their CP-odd components are\footnote{At this order also sub-leadings in 
the CP-even sector contribute. Although we do no illustrate them, all orders 
in $\Delta m^2_{atm}$ are included in the numerical computations.}:
\bea
\label{cpodd1}
\PCPV(\nu_e \to \nu_e) & = & \PCPV(\nu_\mu \to \nu_\mu)  = 
                             \PCPV(\nu_\tau \to \nu_\tau) = 0 \ , \\
\PCPV(\nu_e \to \nu_\mu) & = & 
     8 c^2_{13} c^2_{23} c_{24} c_{34} s_{24} s_{34} \ 
     \sin (\delta_2 + \delta_3) \
     \left( {{ \Delta m^2_{34} L }\over{4 E_\nu} } \right) \ 
     \sin^2 \left (\frac{\Delta m_{23}^2 L}{4 E_\nu} \right) \ , \\
\PCPV(\nu_e \to \nu_\tau) & = & 4 c_{23} c_{24} 
      \Big \{ 
      2 c_{14} s_{14} c_{23} s_{23} s_{13} s_{24} 
      (s^2_{13} s^2_{14} - c^2_{14}) \ \sin (\delta_2 + \delta_3) \nn \\
& + & 
      c_{14} c_{34} s_{13} s_{14} s_{34} \left[ 
      (s^2_{23} - s^2_{24}) \ \sin \delta_2 + 
      s^2_{23} s^2_{24} \ \sin(\delta_2 + 2\delta_3) \right] \\
& + & 
      c_{14} c_{24} s_{13} s_{14} s_{23} s_{24} 
      (c^2_{34} - s^2_{34}) \ \sin \delta_3  \Big \}
      \left( {{ \Delta m^2_{34} L }\over{4 E_\nu} } \right) \  
      \sin^2 \left( \frac{\Delta m_{23}^2 L}{4 E_\nu} \right) \ , \nn \\
\PCPV(\nu_\mu \to \nu_\tau) & = & 
      8 c^2_{13} c^2_{23} c_{24} c_{34} s_{34} \left [ 
      c_{14} c_{23} s_{13} s_{14} \ \sin \delta_2 \ + 
      c^2_{14} s_{23} s_{24} \ \sin (\delta_2 + \delta_3) \right ] \times \nn \\ 
&   & \qquad \left( {{ \Delta m^2_{34} L }\over{4 E} } \right ) \  
      \sin^2 \left( \frac{\Delta m_{23}^2 L}{4 E} \right ).
\label{cpodd2}
\eea

Two distinct phases appear, $\delta_2$ and $\delta_3$, 
in a typical sinusoidal dependence which is the trademark of CP-violation 
and ensures different transition rates for neutrinos and antineutrinos.

CP-odd effects are observable in ``appearance'' channels, while 
``disappearance'' ones are only sensitive to the CP-even part. The latter 
is mandated by CPT \cite{bernabeu}. In contrast with the three-neutrino
case, the solar suppression (see \cite{dgh}) is now replaced by the a less 
severe atmospheric suppression. CP-violating effects are then expected to 
be one or two orders of magnitude larger that in the standard case, and 
independent of the solar parameters. 

Staying in the ``conservative'' assumption of small $\theta_{13},\theta_{14}, 
\theta_{23}, \theta_{24}$, we compare two democratic scenarios, in which 
all these angles are taken to be small and of the same order:
\begin{enumerate}
\item Set 1: $\theta_{34} = 45^\circ$, $\theta_{ij} = 5^\circ$ 
      and $\Delta m^2_{atm} = 2.8 \times 10^{-3}$ eV$^2$
      for $\Delta m^2_{LSND} = 0.3$ eV$^2$;
\item Set 2: $\theta_{34} = 45^\circ$, $\theta_{ij} = 2^\circ$ and 
      $\Delta m^2_{atm} = 2.8 \times 10^{-3}$ eV$^2$ for 
      $\Delta m^2_{LSND} = 1$ eV$^2$.
\end{enumerate}
The value chosen for $\Delta m^2_{atm} $ is the central 
one of the most recent SuperK analysis \cite{Superka}. In the figures below, 
the exact formulae for the probabilities have been used.

The size of the CP-asymmetries is very different for $\mu$ channels and 
$\tau$ channels. For instance for Set 2, they turn out to be small 
in $\nu_e-\nu_\mu$ oscillations, ranging from the permil level to a few percent. 
In contrast, in $\nu_\mu-\nu_\tau$ oscillations they attain much larger values 
of about $50\%-90\%$. This means that their hypotehtical measurement should be 
rather insensitive to systematic effects, and other conventional neutrino beams 
from pion and  kaon decay could be appropiate for their study.

%
\begin{figure}[t]
\vspace{0.1cm}
\begin{tabular}{cc}
\epsfig{figure=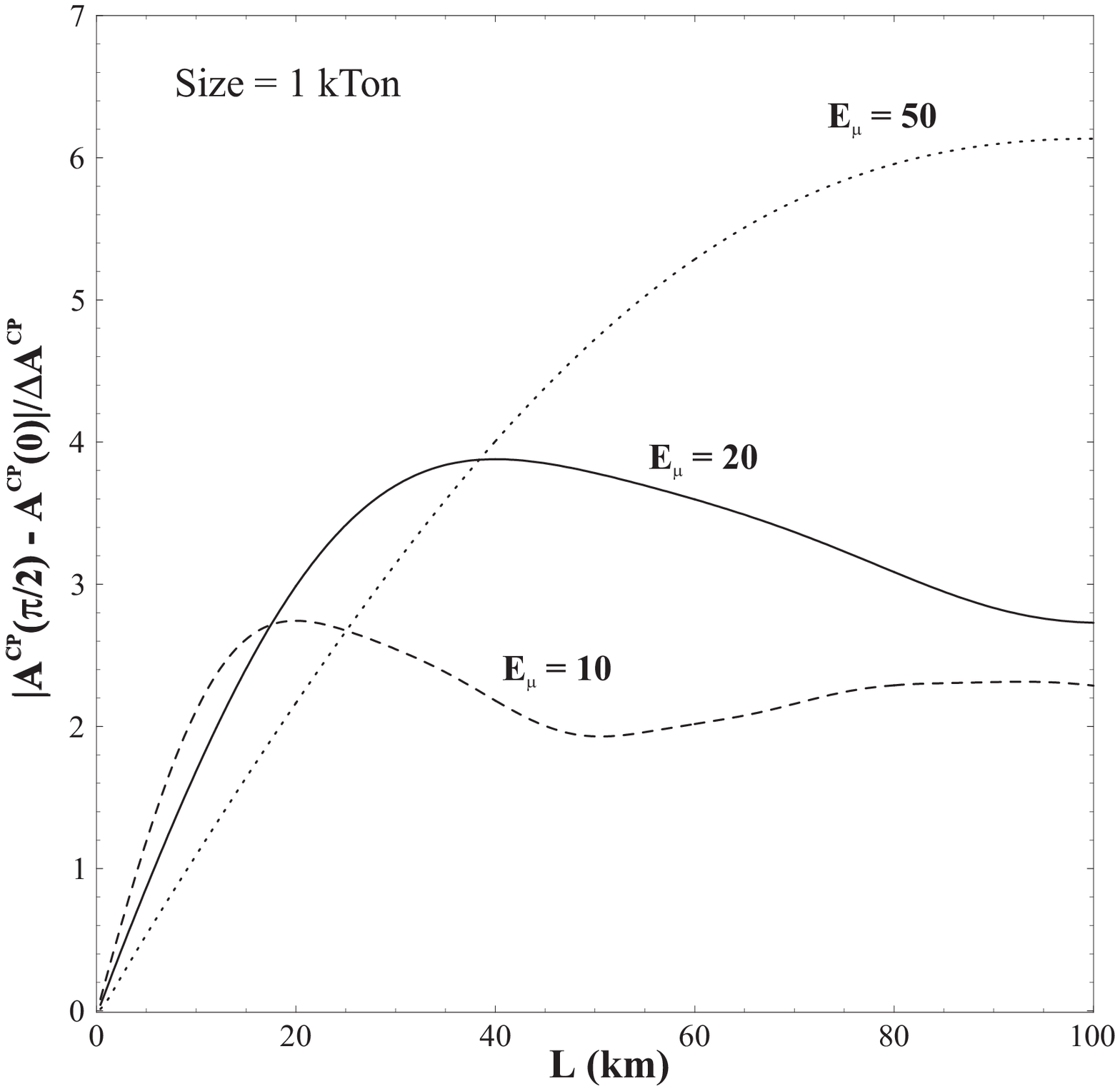,height=7.4cm,angle=0} & 
\epsfig{figure=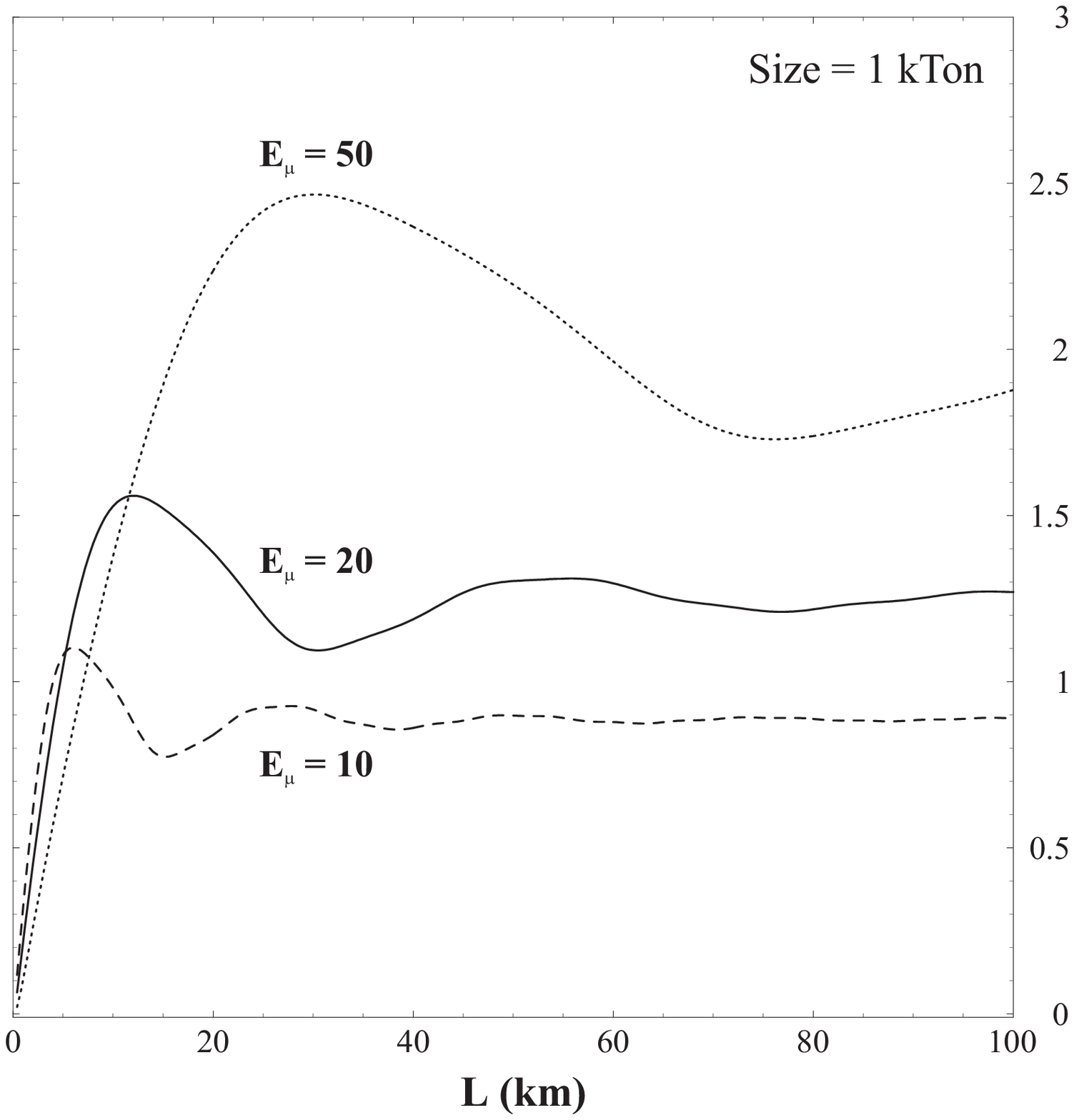,height=7.4cm,angle=0} 
\end{tabular}
\caption{\it{
Signal over statistical uncertainty for CP violation, in the 
$\nu_e \to \nu_\mu$ channel, for the two sets of parameters described in the 
text (Set 1 on the left and Set 2 on the right). We consider a $1$ kTon 
detector and $2 \times 10^{20}$ useful muons/year.}} 
\label{CPetfig12}
\end{figure}
%
%
\begin{figure}[t]
\vspace{0.1cm}
\begin{tabular}{cc}
\epsfig{figure=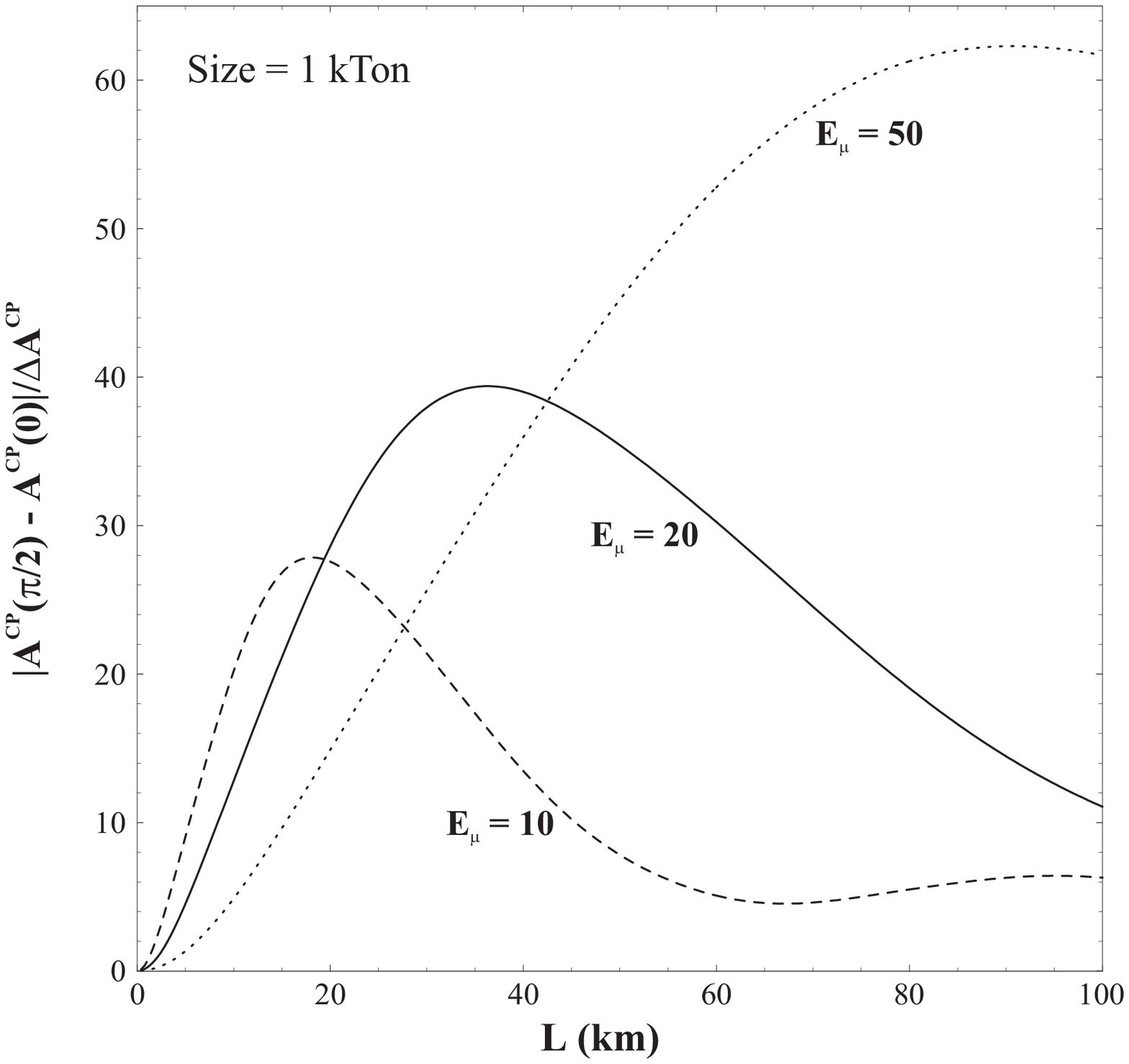,height=7.4cm,angle=0} & 
\epsfig{figure=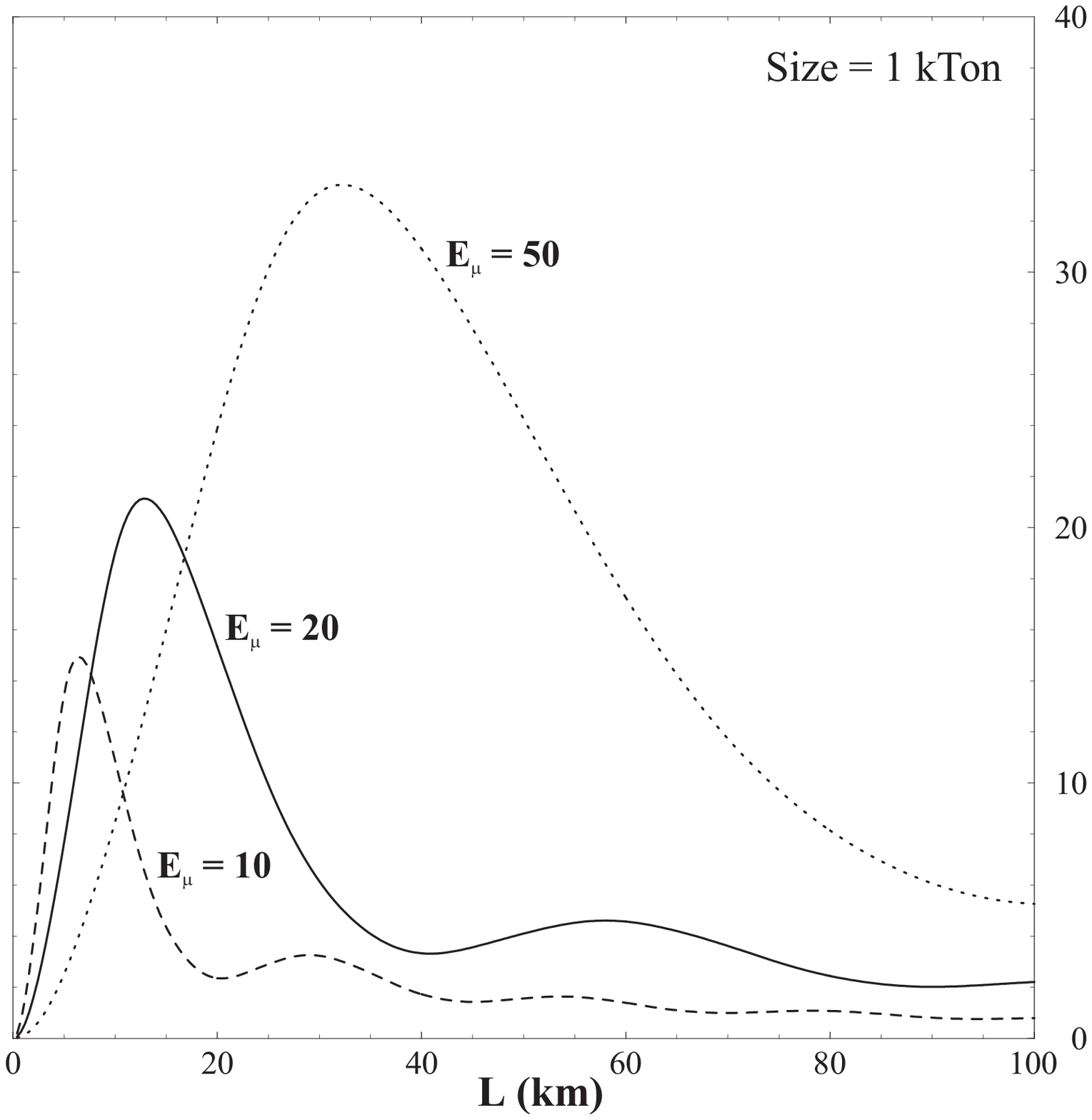,height=7.4cm,angle=0} 
\end{tabular}
\caption{\it{
Signal over statistical uncertainty for CP violation,
in the $\nu_\mu \to \nu_\tau$ channel, for the two sets of parameters described 
in the text (Set 1 on the left and Set 2 on the right). We consider a $1$ kTon 
detector and $2 \times 10^{20}$ useful muons/year.}} 
\label{CPmtfig12}
\end{figure}
%
%
\begin{itemize}
\item {\bf $\mu$ appearance channels}
\end{itemize}
Fig.~\ref{CPetfig12} shows the signal over noise ratio for the integrated 
CP asymmetry, eqs. (\ref{intasy}), in the wrong 
sign muon channel, that is $\nu_e \to 
\nu_\mu$ versus $\bar\nu_e \to \bar\nu_\mu$ oscillations, as a function of 
the distance. Matter effects, although negligible, have been included.
For the scenario and distances discussed here, the scaling laws are analogous to those 
derived for three neutrino species in vacuum, eq. (\ref{scaling}), that is
\be
\frac{A_{e \mu}^{CP}}{\Delta A_{e \mu}^{CP}} \propto \sqrt{E_\nu}  
\left| \sin \left ( \frac{\Delta m^2_{34} \, L}{4 E_\nu} \right ) \right| .
\label{scaling2}
\ee
The maxima of the curves move towards larger distances when the energy of the 
muon beam is increased, or the assumed LSND mass difference is decreased. 
Moreover, increasing the energy enhances the significance of the effect at the 
maximum as expected. 
At $E_\mu = 50$ GeV, 6 standard deviation (sd) signals are attainable at 
around 100 km for the values in Set 1, and just 2.5 sd at 30 km for Set 2, 
levelling off at larger distances and finally diminishing when $E_\nu/L$ 
approaches the atmospheric range.

\begin{itemize}
\item {\bf $\tau$ appearance channels}
\end{itemize}
In Fig.~\ref{CPmtfig12} we show the signal over noise ratio in $\nu_\mu 
\to \nu_\tau$ versus $\bar\nu_\mu \to \bar\nu_\tau$ oscillations as a 
function of the distance. The experimental asymmetry is obtained from 
eq.~(\ref{intasy}), with the obvious replacements $e \to \mu$ and $\mu \to\tau$.
A larger enhancement takes place in this channel as compared to the 
$\nu_e \to \nu_\mu$ one. over 60 sd for Set 1 and 33 sd for Set 2 are a 
priori attainable. These larger factors follow from the fact that the CP-even 
transition probability $P_{CP}(\nu_\mu \nu_\tau)$ is smaller than 
$P_{CP}(\nu_e \nu_\mu)$, due to a  stronger suppression in small mixing angles.
Notice that the opposite happens in the 3-species case. 
Bilenky {\em et al.} \cite{giunti} had previously concluded that 
the $\tau$ channel was best for CP-studies in the four species scenario. 
Their argument relied, though, on the fact that the parameter space involved 
in $\nu_\tau$ oscillations  is experimentally less constrained than the 
$\nu_{\mu}$ one, a freedom we have not used here, staying within the more 
natural assumption that all the angles in the next-to-minimal scheme,
except the atmospheric one, $\theta_{34}$, are small.

The results in the $\nu_e \to \nu_\tau$ channels are almost identical   
to the $\nu_e \to \nu_\mu$ ones, not deserving a separate discussion.

%
%
\begin{figure}[t]
\vspace{0.1cm}
\begin{tabular}{cc}
\epsfig{figure=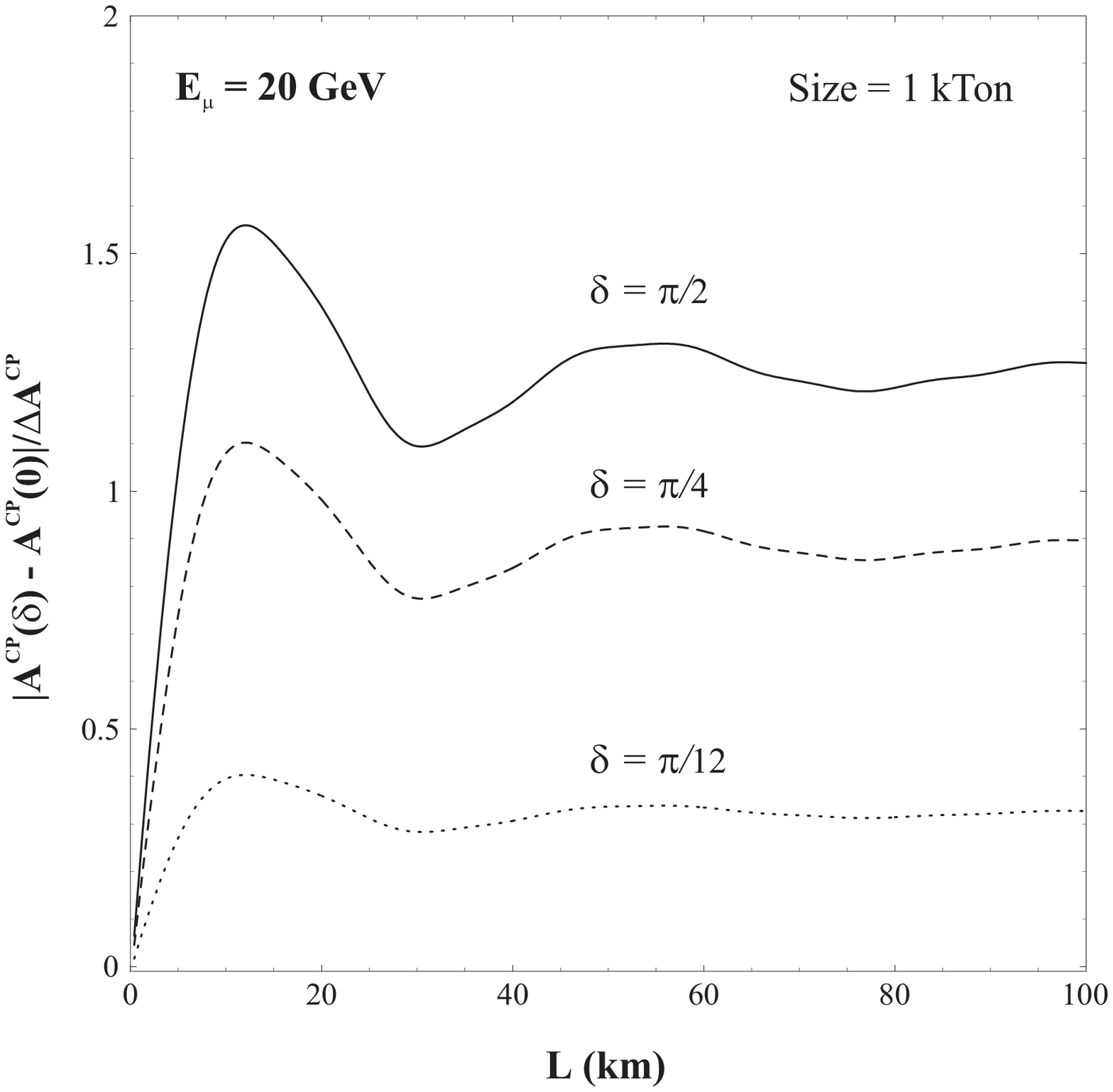,height=7.4cm,angle=0} & 
\epsfig{figure=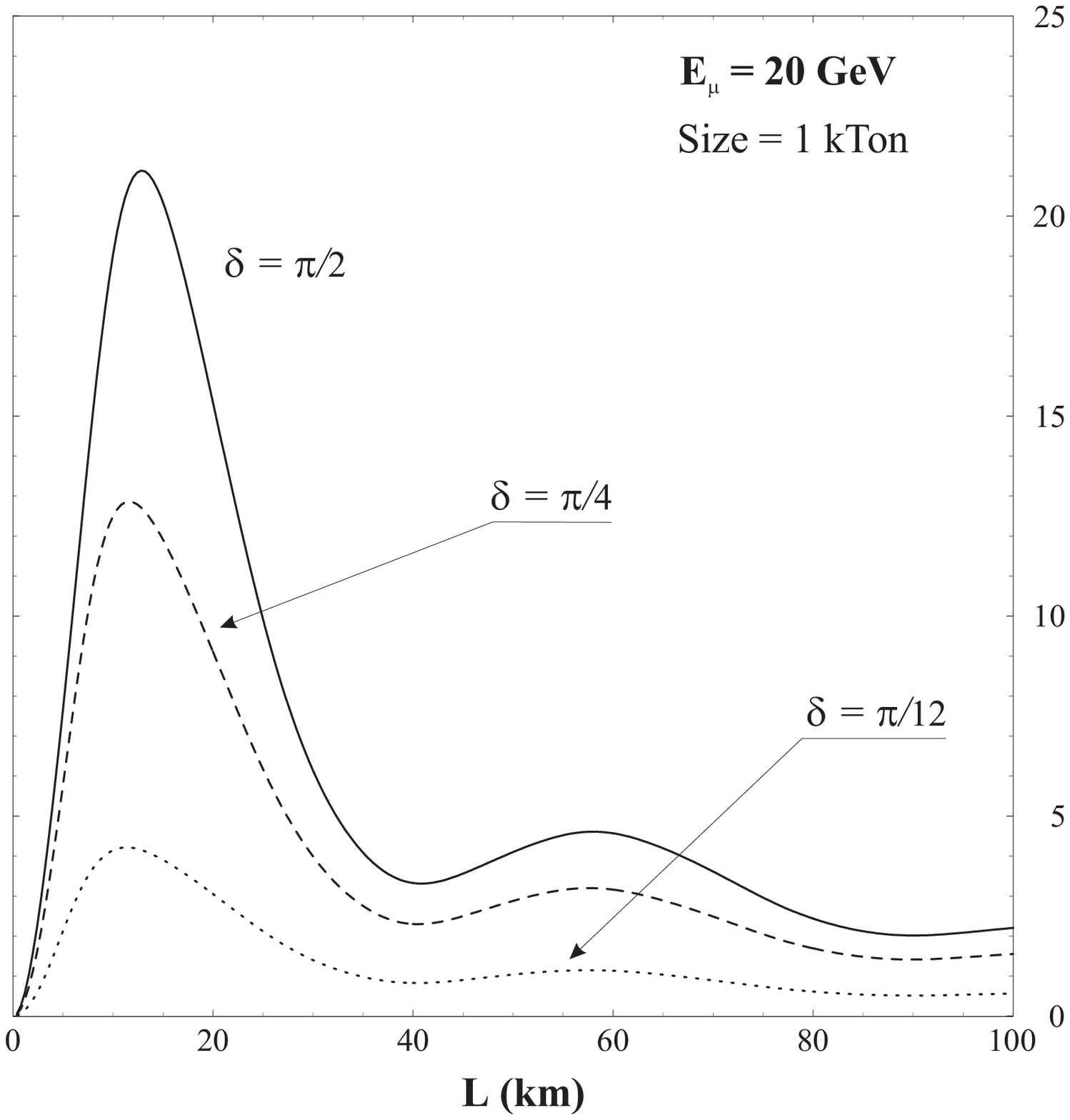,height=7.4cm,angle=0} 
\end{tabular}
\caption{\it
CP violation asymmetry in the $\nu_e \to \nu_\mu$ (left) and 
$\nu_\mu \to \nu_\tau$ (right) channel for $E_\mu=20$ GeV, angles and mass 
differences as in Set 2 and for different choice of the CP phases: 
$\delta_1=\delta_2=\delta_3=\pi/2$ (full line), $\pi/4$ (dashed line) and 
$\pi/12$ (dotted line). We consider a $1$ kTon detector from the 
source of a $2 \times 10^{20}$ muon/year beam.} 
\label{CPphfig}
\end{figure}
%
%

The phase dependence is shown in Fig.~\ref{CPphfig}, with the expected 
depletion of the signal for small CP phases. For small values of the phases, 
i.e. $\delta_1=\delta_2=\delta_3 = 15^\circ$, the significance drops to 
the $1 \sigma$ level. 

%
\section{Summary.}
%
%
The ensemble of solar plus atmospheric neutrino data, when analysed in a 
three-generation mixing scenario, points out the importance of long baseline 
experiments searching for $\nu_e \leftrightarrow \nu_\mu$ transitions. This is 
particularly relevant for determining $\theta_{13}$ and CP-odd effects. 
A neutrino factory from muon storage beams has much higher 
precision and discovery potential than any other planned facility. 
The number of useful observables is sufficient to determine or very 
significantly constrain the parameters $\theta_{23}$ and $\theta_{13}$ and 
$\Delta m^2_{23}$ of a standard three-generation mixing scheme. 
Experiments searching for the appearance of ``wrong
sign'' muons are very sensitive to the parameter space.
For instance, while all other planned experiments will 
reach at most sensitivities of $\sin^2(\theta_{13})> 10^{-2}$, much lower 
values are attainable at the neutrino factory, $\sin^2(\theta_{13})>10^{-4}$.
For CP-violation we have updated previous studies, based in 
``$\mu$ appearance'' channels, which are the most promising ones, and derived 
the scaling laws with distance 
and energy of CP-odd observables. A fair chance to observe a significant 
signal of CP-violation requires that nature chooses $\Delta m^2_{solar}$ in 
the higher range allowed by the ensemble of solar experiments, 
$\Delta m^2_{solar}\sim 10^{-4}$ eV$^2$ and that the solar mixing is 
large\footnote{If the results of some solar neutrino experiment are disregarded 
allowing for larger value for the solar mass difference, i.e. 
$\Delta m^2_{solar}= 8\,10^{-4} eV^2$, a promising signal follows even
with moderate neutrino fluxes, as discussed in the text.}. 
In other words, it requires that the large mixing angle 
solution (LMA-MSW) of the  solar deficit is confirmed by the solar 
experiments. Would that be the case, a neutrino factory with 
high intensity muon beams ($\sim 10^{21}$ muons decaying in the direction 
of the detector) could be precise enough to discover CP-violation in the 
lepton sector. Matter effects in the energy integrated CP-odd observables 
that we have considered get more easily disentagled from the truly 
CP-violating effects the smaller $\theta_{13}$ is.

When the LSND signal is also taken into account, the reach of 
short base line experiments is extremely large.
We have derived one and two mass scale dominance approximations, 
appropriate for CP-even and CP-odd observables, respectively.
The plethora of channels for different flavours provided by the 
neutrino factory allows to cover the full mixing parameter space.
CP violation may then be easily at reach, specially through 
``$\tau$ appearance'' signals. In these channels the CP-asymmetries are 
so large that even neutrino beams from conventional pion and kaon decays 
may be sufficient for their detection.     

\section{Acknowledgements}
We acknowledge useful conversations with:
B. Autin, R. Barbieri, J. Bernabeu, S. Bilenky,  C. Giunti, A. De R\'ujula, 
F. Dydak, J. Ellis, J. G\'omez-Cadenas, 
M.C. Gonzalez-Garcia, O. Mena, S. Petcov, C. Quigg and A. Romanino. 
A. D., M. B. G., and S. R. thank the CERN Theory Division for hospitality 
during the final stage of this work; their work was also partially supported 
as well by CICYT project AEN/97/1678. A. Donini acknowledges the I.N.F.N. 
for financial support. S. Rigolin acknowledges the European Union for 
financial support through contract ERBFMBICT972474.

\end{document}